\documentclass[prd,preprintnumbers,superscriptaddress,nofootinbib,floatfix,twocolumn,notitlepage]{revtex4}
\usepackage[dvips]{graphicx}
\usepackage{dcolumn} 
\usepackage{bm}
\usepackage{epsfig,amsmath,amssymb,verbatim,mathrsfs,array,layout,textcomp,amssymb,latexsym,slashed}
\usepackage{color}
\usepackage[colorlinks=true,citecolor=blue,urlcolor=blue,linktocpage=true,
linkcolor=blue]{hyperref}
\usepackage[utf8]{inputenc}
\usepackage{multirow}

\input epsf.tex
\newcommand{\beq}{\begin{eqnarray}}
\newcommand{\eeq}{\end{eqnarray}}

\def\ltap{\ \raise.3ex\hbox{$<$\kern-.75em\lower1ex\hbox{$\sim$}}\ }
\def\gtap{\ \raise.3ex\hbox{$>$\kern-.75em\lower1ex\hbox{$\sim$}}\ }

\def\be{\begin{equation}}
\def\ee{\end{equation}}
\def\bea{\begin{eqnarray}}
\def\eea{\end{eqnarray}}

\pdfoutput=1

\definecolor{newred}{rgb}{0.5,0.1,0}
\definecolor{darkgreen}{rgb}{0.0,0.7,0.2}
\definecolor{lightblue}{rgb}{0.0,0.5,1}

\RequirePackage[normalem]{ulem} 
\RequirePackage{color}\definecolor{RED}{rgb}{1,0,0}\definecolor{BLUE}{rgb}{0,0,1} 

\begin{document} 

\title{Dark Matter Abundance from Sequential Freeze-in Mechanism}
\preprint{LAPTH-021/20}
\preprint{IFT-UAM/CSIC-20-65}

\author{Genevi\`eve B\'elanger}
\affiliation{Laboratoire d'Annecy-le-Vieux de Physique Th\'eorique LAPTh, CNRS -- USMB, BP 110 Annecy-le-Vieux, F-74941 Annecy, France}
\author{C\'edric Delaunay}
\affiliation{Laboratoire d'Annecy-le-Vieux de Physique Th\'eorique LAPTh, CNRS -- USMB, BP 110 Annecy-le-Vieux, F-74941 Annecy, France}
\author{Alexander Pukhov}
\affiliation{Skobeltsyn Institute of Nuclear Physics, Moscow State University, Moscow 119992, Russia}
\author{Bryan Zaldivar}
\affiliation{Departamento de Fisica Teorica and Instituto de Fisica Teorica, IFT-UAM/CSIC, Cantoblanco, 28049, Madrid, Spain}

\begin{abstract}
We present a thorough analysis of the sequential freeze-in mechanism for dark matter production in the early universe. In this mechanism the dark matter relic density results from pair annihilation of mediator particles which are themselves produced by thermal collisions of standard model particles. Below some critical value of the mediator coupling to standard model fields, this sequential channel dominates over the usual freeze-in where dark matter is directly produced from thermal collisions, even when the mediator is not in thermal equilibrium. The latter case requires computing the full non-thermal distribution of the mediators, for which finite temperature corrections are particularly important.
\end{abstract}

\maketitle

\section{Introduction}
The nature of the dark matter (DM) is perhaps the most acute open question in particle physics. Despite the strong observational evidence for an important DM component in the Universe, most of its properties remains unknown. Requiring that the DM be a thermal relic implies non gravitational interactions with ordinary matter. This nourishes hope to probe the DM in our local environment, either by detecting DM particles, directly in underground experiments or indirectly through the products of DM annihilation within our galactic neighborhood, or by producing them at colliders. 

In particular, the hypothesis of a new DM particle around the electroweak scale has been tested extensively and the lack of evidence for DM from these searches triggered a growing interest in exploring a wider class of DM models.  One possibility is that the DM and/or the mediator connecting it to the SM is below the GeV scale, thus leaving no traces in nuclear-scattering based direct detection experiments and colliders. This regime can be tested nonetheless with different experimental probes~\cite{Alexander:2016aln,Battaglieri:2017aum}. Besides, a mediator in the sub-GeV mass range also helps in resolving small-scale problems related to DM~\cite{Spergel:1999mh,BoylanKolchin:2011de,Oman:2015xda}.

Another possibility is that the DM particle still lies above the GeV scale but signals in standard searches are suppressed because it interacts only very weakly with the SM. In this scenario, the DM is never in equilibrium with the SM in the early Universe and is instead produced by freeze-in through pair annihilation or decay of particles in the thermal bath~\cite{McDonald:2001vt,Hall:2009bx}. The DM is generally assumed to be singlet under the SM group and part of a hidden sector that couples to the SM through renormalizable portal operators or with a mediator particle. Despite its tiny couplings with the SM fields the DM relic density often remains detectable in existing experiments~\cite{Bernal:2017kxu}, like direct detection when the mediator is light~\cite{Hambye:2018dpi,Essig:2017kqs} or indirect detection for decaying DM~\cite{Arcadi:2013aba,Roland:2015yoa}. Frozen-in DM could also be tested in  cosmology~\cite{Fradette:2014sza,Bernal:2015ova,Berger:2016vxi} and at colliders through signatures involving displaced vertices and/or long-lived particles~\cite{Co:2015pka,Evans:2016zau,Hessler:2016kwm,Ghosh:2017vhe,Calibbi:2018fqf,Belanger:2018sti,Alimena:2019zri} including with detectors located some distance from the interaction point~\cite{Curtin:2018mvb,Ariga:2018uku,No:2019gvl,Heeba:2019jho}.\\

In this work we consider scenarios where dark sector particles are feebly coupled to the SM and where the DM relic is produced non-thermally through the freeze-in mechanism. We assume the mediator mass is small, at the $10\,$MeV scale, and  is a scalar, for simplicity.\footnote{The case of a light vector is qualitatively similar, see {\it e.g.} Refs.~\cite{Boehm:2003hm, Zurek-lightDM}.} Lighter mediators also have  interesting phenomenology but suffer severe constraints from Big Bang nucleosynthesis.  The light mediator can potentially  provide an explanation for anomalies observed at the cluster scale~\cite{Buckley:2009in,Kaplinghat:2015aga} and enhances the DM-nucleus scattering cross section, which offers the possibility to probe this scenario in direct detection. We will consider a simplified  model with an hadrophilic scalar mediator which couples only to light quarks, thus alleviating several constraints that affect Higgs portal models where couplings to heavy quarks dominate ~\cite{Kim:2008pp,LopezHonorez:2012kv,Krnjaic:2015mbs}.  This framework will be sufficient to  illustrate the main phenomenological properties   that are  expected to be relevant for a larger set of models.

For freeze-in to take place, the product of mediator couplings to the SM and to DM must be very small, ${\cal O}(10^{-12} - 10^{-10})$, while their relative size remains a free parameter~\cite{Hall:2009bx}. In models where DM is much heavier than the mediator, DM can generically be produced via pair annihilation of SM particles or mediators, assuming the latter are in equilibrium with the SM thermal bath. 
 Here, we point out that even when the mediator coupling to the SM is too small for the mediator to ever reach equilibrium,  a finite density of mediators can be produced through SM induced processes.
The contribution of such non-thermal mediators to DM production parametrically dominates over that of pair annihilation of SM particles. This new phase of DM production, called sequential freeze-in~\cite{Hambye:2019dwd},  requires first solving for the momentum distribution of the mediator yield before using it for DM production. 
For this we solved the unintegrated Boltzmann equation rather than assuming $\phi$ to be in kinetic equilibrium with the thermal bath.
 Because the mediator is much lighter than DM, the tail of the mediator distribution is most relevant for DM production.  Moreover we show that thermal effects, which we  approximate by taking into account thermal masses for quarks, gluons and photons
\footnote{A recent study showed the importance of the thermal mass for DM production from thermal photon decays~\cite{Dvorkin:2019zdi}. } play an important role in mediator production. Finally, our calculation takes into account the Bose-Einstein and Fermi-Dirac distributions for bosons and fermions, respectively, rather than simply using Maxwell-Boltzmann (MB) distributions. We show that these effects lead to large corrections in the calculation of the relic density~\cite{Belanger:2018mqt}.\\
 
The paper is organised as follows. We first present the simplified model and discuss at length the possible DM production mechanisms, giving an extensive description of the specific case where the mediator is not in thermal equilibrium (Section 3).  With the complete calculation of DM production in hand (Section 4), we then 
determine the potential of current and future DD detectors to probe this model (Section 5) and examine numerous astrophysics and terrestrial constraints (Section 6). Our conclusions are presented in Section 7.  The appendices contain details on the reaction rates for mediator production as well as an approximate analytical solution for the mediator distribution.

\section{Simplified dark matter model} 
\label{sec:model}

Consider a simplified model for the dark sector which comprises a Dirac fermion $\chi$ (the DM candidate) interacting with SM quarks through a real scalar mediator $\phi$ with parity-preserving renormalizable couplings
\beq\label{Lint}
-\mathcal{L}_{\rm int}= y_\chi \phi\bar\chi\chi + y_q \phi\bar q q\,,
\eeq  
where the Lagrangian above is written below the scale of EW symmetry breaking. 
Both dark states are neutral under the SM gauge group and $\chi$ is assumed odd under a $\mathbb{Z}_2$ symmetry and is thus stable. In the following, we will only consider a nonzero coupling for the first-generation up quark, $q=u$.\footnote{A similar model was considered in ~\cite{Batell:2018fqo}  and  flavour issues in this class of models were addressed in ~\cite{Batell:2017kty,Egana-Ugrinovic:2018znw}.} 
 
The question of the origin of the interactions in Eq.~\eqref{Lint} might be raised. If $\phi$ is a SM singlet, we envisage two simple possibilities. For instance, $\phi$ and the SM could connect directly through the (renormalizable) Higgs portal, giving $y_f=m_f/v\times \sin\theta$ where $v\approx246\,$GeV is the SM Higgs VEV breaking EW symmetry and $\theta$ is a mixing angle. In this case though, the mediator would couple to all charged fermions and its interactions with the third family would dominate.\footnote{See Ref.~\cite{Krnjaic:2015mbs} for a detailed study of thermal DM production in this scenario.} In the absence of Higgs portal, $\phi$ could couple to SM fermions through interactions that involve additionnal states beyond the SM. For example, SM quarks could mix with new vector-like fermions that undergo Yukawa interactions with $\phi$. 
In the limit that these additional states are heavy, characterized by a mass scale $\Lambda$, their dynamics will be captured by non-renormalizable operators like $(x_f/\Lambda)\bar QHf_R\,\phi+{\rm h.c.}$, with $Q$ and $f=u,d$ denoting respectively SU(2)$_L$ doublets and singlets. 
 In this case $y_f\sim x_fv/\Lambda$ and coupling to the first family could dominate depending on the UV dynamics setting the flavor structure of the $x_f$ couplings. Moreover, taking $\Lambda\gg v$ would provide a simple rationale for the feeble couplings characterizing the freeze-in mechanism which we ought to consider in this article.\\ 

At energies below $\Lambda_{\rm QCD}\approx 200\,$MeV, quarks are no longer free and the $\phi$ interactions with the SM is better described in terms of hadronic resonances. We will limit ourselves here to protons, neutrons and pions. The low-energy interaction Lagrangian is 
\beq
-\mathcal{L}_{\rm int}^{\rm had}=\sum_{N=p,n} y_N \phi \bar N N+ y_\pi \phi \pi \pi\,,
\eeq
where the hadronic couplings are related to the fundamental quark coupling by matching. For first generation quarks, the coupling to nucleons is 
\beq
\label{eq:yp}
y_N=y_q\frac{m_N}{m_q} f_q^N\,,
\eeq
where the $f_q^N$'s are nuclear form factors whose values are extracted from matching nuclear data with lattice simulations~\cite{Shifman:1978zn,Bhattacharya:2016zcn}. For energies well below $4\pi f_\pi$, where $f_\pi\approx 93\,$MeV is the pion decay constant, the coupling to pions can be derived from chiral perturbation theory, which gives 
\beq
y_\pi =  y_q\frac{m_\pi^2}{m_u+m_d}\,,
\eeq
at leading order for first generation quarks~\cite{Donoghue:1990xh,Bijnens:1998fm}.\\ 

In order to retain the possibility of velocity-dependent cross sections for DM self-interactions required by clusters anomaly, we will consider the light mediator limit $m_\phi\ll m_\chi$ with $m_\phi >1\,$MeV.  For mediators below the MeV scale, very strong constraints apply on scalar couplings to nucleons which basically excludes freeze-in production of DM~\cite{Zurek-lightDM}.  

Here the mediator is $\mathbb{Z}_2$-even and thus unstable. For $m_\phi< 2m_\pi$ the leading decay channel is in two photons through loops of charged hadrons. 
\begin{equation}\label{phiwidth}
\Gamma_{\phi\to\gamma\gamma}= \frac{\alpha^2 m_\phi^3}{144 \pi^3} \left|\frac{y_N}{m_N}+\frac{y_\pi}{8m_\pi^2}\right|^2\,,
\end{equation}
which corresponds to a lifetime of  $\tau_\phi\approx 8.4\times 10^{-11}\,$sec$/y_q^2$ for $m_\phi=1\,$MeV and a $q=u$ coupling evaluated at the matching scale $\mu=2\,$GeV. The decay of such a light, long-lived mediator would typically alter big bang nucleosynthesis (BBN), unless the mediator decouples and decays before it starts. At $m_\phi=1\,$MeV, the model is in tension with BBN for $y_q\lesssim \mathcal{O}(10^{-5})$, greatly restricting the region of parameter space favored by the freeze-in mechanism. This strong constraint can be easily evaded by, for instance, introducing in $\mathcal{L}_{\rm int}$ an additional interaction of the mediator to neutrinos, $-y_\nu \phi\bar\nu\nu$. The decay width into neutrinos $\Gamma_{\phi\to \nu\bar\nu}= y_\nu^2 m_\phi/8\pi$ can be sufficiently large to avoid  BBN constraints with a relatively small coupling $y_\nu \sim\mathcal{O}(10^{-10})$ which has no significant impact on the DM phenomenology. Note that the mediator has to decay into neutrinos sufficiently early  so that most of them thermalize before neutrino decoupling at $T\approx {\rm few}\,$MeV. 

In order to avoid strong constraints from BBN while maximizing the effect of the light mediator on DM phenomenology we conservatively set  $m_\phi=10\,$MeV in the remainder of this article. 
 Moreover, we focus on DM in the $1-100\,$GeV mass range where significant DM-nucleus scattering signals in next-generation direct detection experiments are expected. \\

\section{Mediator freeze-in production}
\label{sec:phireezein}

The mediator contribution to DM production $\gamma_{\phi\phi\to\chi\bar\chi}$ requires knownledge of the phase-space distribution of $\phi$ particles, $f_\phi(p,T)$. The latter is obtained from solving the (unintegrated) Boltzmann equation 
\beq\label{Beqphi}
E(\partial_t-Hp\partial_p)f_\phi = C[f_\phi]\,,
\eeq
where $p$ and $E=(m_\phi^2+p^2)^{1/2}$ are respectively the 3-momentum and energy of $\phi$ in the frame of the thermal bath, $H$ is the Hubble rate  and $C[f_\phi]$ is the collision term. Solving Eq.~\eqref{Beqphi} is numerically challenging partly because of the $\partial_p$ term which account for the momentum change due Hubble expansion. It is however possible to factor out this effect by introducing the dimensionless variable 
\begin{equation}\label{qvar}
q\equiv \frac{p}{T_0}\left[\frac{s(T_0)}{s(T)}\right]^{1/3}=\frac{p}{T}\left[ \frac{h_{\rm eff}(T_0)}{h_{\rm eff}(T)}\right]^{1/3}
\end{equation}
where $h_{\rm eff}(T)$ is the number of degrees of freedom contributing to the entropy density $s(T)=2\pi^2/45 h_{\rm eff} T^3$ and $T_0\approx2.3\times10^{-4}\,$eV is the photon temperature today. The entropy ratio in Eq.~\eqref{qvar} further accounts for the slowdown of the Hubble rate due to the decoupling of species across the cosmic history. For $T\gtrsim\mathcal{O}(1\,$GeV), $q\approx 0.4\times p/T$.

In terms of this reduced momentum variable 
 Eq.~\eqref{Beqphi} is brought to a one-derivative differential equation
\beq\label{Beqphi2}
x\bar H\partial_x f_\phi(q,x)=E^{-1}C[f_\phi]\,,
\eeq
here expressed in terms of $x=m_\chi/T$, with $\bar H\equiv H/[1+1/3\times d\log h_{\rm eff}/d\log T]$. This equation can be solved for fixed $q$.

The other complication lies in the form of the collision term $C[f_\phi]$. Several interactions contribute to bring the mediator in thermal contact with the SM bath. The dominant contribution arises from QCD processes with a single $\phi$ in the final state. 
Those are $gq\to q\phi$ and $\bar qq\to g\phi$ as well as $g\to q\bar q\phi$ once thermal corrections are included (see section~\ref{sec:temperature}). We also include subdominant electromagnetic processes $q\gamma\to q\phi$ and $q\bar q\to \gamma\phi$ which contributes at the $\mathcal{O}(10\%)$ level. Pair production of $\phi$ is suppressed by a factor of $\mathcal{O}(4\pi y_q^2/\alpha_s)$, where $\alpha_s$ is the QCD coupling, and is therefore negligible in the limit $y_q\ll 1$ that is required for freeze-in. Hence, we have 
\beq\label{RHSphiBE}
E^{-1}C[f_\phi] = \hat\gamma_{q\bar q\leftrightarrow V\phi}+2\hat\gamma_{qV\leftrightarrow q\phi} + \hat\gamma_{g\leftrightarrow q\bar q\phi}\,,
\eeq
where $\hat\gamma_i\equiv E^{-1}C_i[f_\phi]$ is the collision term associated with the process $i$. A sum over $V=g,\gamma$ is implicit in Eq.~\eqref{RHSphiBE} and the factor 2 multiplying the second term on the right-hand side accounts for the charge-conjugated process $\bar q V\leftrightarrow \bar q\phi$. Note that the last term is only sourced by thermal plasma effects.\\

\subsection{Unintegrated collision rates}
The collision terms for $2\to 2$ and $1\to 3$ processes are generally expressed as (upper/lower sign applies to fermions/bosons)
\begin{widetext}
\beq\label{collterm}
\hat\gamma_{{\rm in}\leftrightarrow {\rm out}+\phi} = \frac{1}{2E}\int \prod_{i}\frac{d^3p_i}{(2\pi)^32E_i}(2\pi)^4\delta^{(4)}(P_{\rm in}-P_{\rm out}-p)\left[f_{\rm in}(1\mp f_{\rm out})(1+ f_\phi)|\mathcal{M}_{{\rm in}\to {\rm out}+\phi}|^2 - ({\rm in}\leftrightarrow {\rm out}+\phi)\right]\,,
\eeq
\end{widetext}
where $\mathcal{M}$ denotes the scattering amplitude with initial {\rm and} final state spins and colors summed over, the in (out) label denotes all the incoming (outgoing) particles other than $\phi$ with total momentum $P_{\rm in(out)}$, the index $i$ runs over bath particles and $f_{\rm in}(1\mp f_{\rm out})=f_1f_2(1\mp f_3)$ for $2\to 2$ scattering and $f_1(1\mp f_2)(1\mp f_3)$ for $1\to 3$ decay processes. The first term inside the bracket represents the creation contribution from the process ${\rm in}\to {\rm out}+\phi$, while the second one accounts for depletion from its reverse counterpart ${\rm out}+\phi\to {\rm in}$. 

These two contributions, respectively denoted by $\hat\gamma_{{\rm in}\to{\rm out}+\phi}$ and $\hat\gamma_{{\rm out}+\phi\to {\rm in}}$, are related thanks to equilibrium conditions. Indeed,
 since particles 1, 2 and 3 are in thermal equilibrium with distribution $f_i=[\exp(E_i/T)\pm1]^{-1}$, energy conservation, $E_{\rm in} = E_{\rm out} + E$, implies    
\beq
f_{\rm out}(1\mp f_{\rm in}) = e^{E/T}f_{\rm in}(1\mp f_{\rm out})\,
\eeq
where $E$ is the energy of $\phi$ in the rest frame of the plasma. 
Moreover, in the absence of CP violation (as in the simplified model of interest) $\mathcal{M}_{{\rm in}\to{\rm out}+\phi} = \mathcal{M}_{{\rm out}+\phi\to {\rm in}}$, hence we have 
\beq
\hat\gamma_{{\rm out}+\phi\to {\rm in}} = \frac{e^{E/T}f_\phi}{1+f_\phi}\, \hat\gamma_{{\rm in}\to {\rm out}+\phi}\,.
\eeq 
Then, the generic expression for the collision terms in Eq.~\eqref{Beqphi2} simplifies to
\beq\label{collterm2}
\hat\gamma_{{\rm in} \leftrightarrow {\rm out}+\phi}= R(x,q)\left[(1+f_\phi)e^{-E/T}-f_\phi\right]\,,
\eeq
where 
\beq
R(x,q)\equiv f_\phi^{-1}\hat\gamma_{{\rm out}+\phi\to {\rm in}} =\frac{e^{E/T}}{1+f_\phi} \hat\gamma_{{\rm in}\to {\rm out}+\phi}\,,
\eeq
 is the rate of the reaction ${\rm in} \leftrightarrow {\rm out}+\phi$. Note that rates can be evaluated considering either the creation or depletion process of $\phi$, thanks to equilibrium of SM particles. Finally, in the limit where this rate is much faster than the Hubble rate, $R/\bar H\gg 1$, the bracket in Eq.~\eqref{collterm2} goes to zero, meaning that $\phi$ reaches thermal equilibrium with the SM, $f_\phi\to (e^{E/T}-1)^{-1}$.\\   

For $2\to 2$ scattering, it is more convenient to consider depletion processes (with initial state $\phi$) to calculate the associated rates. Indeed, in this case, neglecting Pauli blocking and stimulated emission effects, {\rm ie.} taking $(1\mp f_1)(1\mp f_2)\simeq 1$, $\hat\gamma_{3\phi\to 12}$ admits a simple expression in terms of the scattering cross-section $\sigma_{3\phi\to 12}$ and,
\beq\label{coll2to2}
R_{\rm scat} \simeq \frac{g_3}{(2\pi)^3} \int d^3p_3 f_3\sigma_{3\phi\to 12}v_{\rm M\o l}\,,
\eeq
where $v_{\rm M\o l}$ is the M\o ller velocity and $g_i$ is the number of degrees of freedom of particle $i$. 

On the other hand, considering creation processes (with final state $\phi$) is more convenient for decay. Within the same approximation, the collision term for the $1\to3$ process is expressed in terms of the differential (partial) decay width $d\Gamma_{1\to 23\phi}/d^3p$ in the frame of the thermal bath, yielding
\beq\label{coll1to3}
R_{\rm decay}\simeq g_1e^{E/T}\int d^3p_1 f_1 \frac{d\Gamma_{1\to23\phi}}{d^3p}\,.
\eeq

We refer the interested reader to Appendix~\ref{app:collisionterms} for a fully detailed evaluation of the integrals in Eqs.~\eqref{coll2to2} and~\eqref{coll1to3}. 
Note that integrating $\hat\gamma$ over the $\phi$ phase space yields $\int d^3p/(2\pi)^{3}\, \hat\gamma_{3\phi\to 12} = \langle \sigma_{3\phi\to 12}v\rangle n_3n_\phi$ and $\int d^3p/(2\pi)^{3}\, \hat\gamma_{1\to 23\phi} =\langle \Gamma_{1\to 23\phi}\rangle n_1$, where  $\langle\cdots\rangle$ denotes thermal averaging and $n$ the number density\\

\begin{figure}[t!]
\begin{center}
\includegraphics[width=0.9\columnwidth]{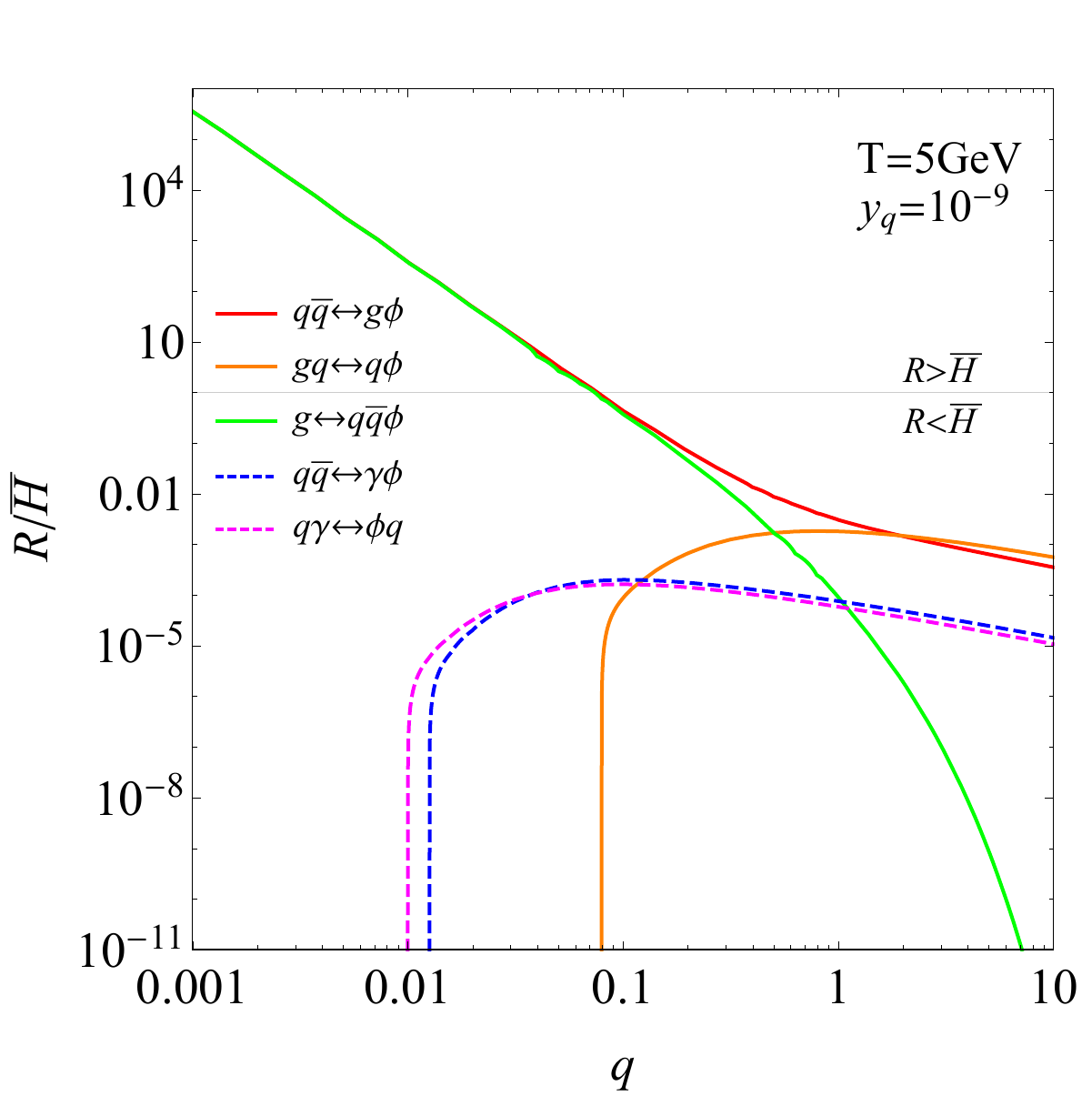}
\caption{Reaction rates relative to the Hubble rate for all processes relevant in $\phi$ production, including thermal masses for SM particles.} 
\label{fig:rates}
\end{center}
\end{figure}

The reaction rates depend on temperature and rescaled momentum $q$, and they are typically faster for low momenta. For instance, as we show in Appendix~\ref{app:collisionterms}, the rate of $2\to 2$ reactions approximately scales like $1/q$ at large $q$. As a result, energetic $\phi$ particles, whose momentum is larger than temperature, take more time to reach equilibrium relative to less energetic ones. For sake of illustration the rates of all relevant processes, including important plasma effects that we discuss in the next section, are shown in Fig.~\ref{fig:rates} for $T=5\,$GeV. Rates for different temperatures show similar behavior (see Appendix~\ref{app:collisionterms}).\\

\subsection{Finite temperature corrections}
\label{sec:temperature}

Thermal corrections to the collision term $C[f_\phi]$ are implemented as follows. The perturbative formulation of gauge theory in vacuum (in powers of gauge coupling) breaks down in the presence of a hot medium due to the emergence of an external scale, the temperature $T$ of the plasma. Gauge theory at finite temperature can still be formulated consistently only with a reorganized perturbative series where a certain class of diagrams needs to be resummed~\cite{Bellac:2011kqa,Kapusta:1989tk} (see also Ref.~\cite{Su:2011zv} for a recent review). We limit ourselves here to the so-called hard-thermal-loop (HTL) approximation~\cite{Braaten:1989mz} which only resums the higher-order loop diagrams associated with soft momenta $\sim gT\lesssim T$ where $g$ denote gauge couplings. In this approximation, fermions and gauge bosons are quasiparticles with temperature dependent masses.

Gauge bosons at finite temperature have polarizarization-dependent dispersion relations~\cite{tsytovich1961spatial}. However, the propagators of  transverse and longitudinal polarizations have the same pole at zero momentum, which is given by the  plasma frequency related to the Debye screening of the (chromo)electric field in a medium, and only develop small differences for non zero momentum. We neglect these differences here and in the calculation of scattering amplitudes we only replace the gauge propagator by a massive one with a pole mass given by the thermal Debye mass. To leading order in gauge coupling those are~\cite{Weldon:1982aq} 
\beq 
m_g^2(T)=g_s^2T^2/3[N_c+n_f(T)/2]\,, 
\eeq
for gluons and 
\beq
m_\gamma^2(T)=e^2T^2n_{\rm ch}(T)/3\,, 
\eeq
for photons, where $n_f(T)$ and $n_{\rm ch}(T)$ are the number of active (namely satisfying $m\lesssim T$) quark flavors and charged particles in the plasma, respectively.

Quark dispersion relations are also modified at finite temperature with the emergence of hole excitations~\cite{Weldon:1989ys}. Nevertheless, particle and hole states are together well described by a quark propagator with a momentum independent thermal mass~\cite{Giudice:2003jh}. We therefore neglect these differences and simply add to the quark propagator the thermal mass~\cite{Weldon:1982bn}, 
\beq
m_q^2(T)=g_s^2T^2/6+e^2T^2Q_q^2/8\,,
\eeq
where $Q_q$ is the quark electric charge, in the calculation of scattering amplitudes. 

Finally, interaction vertices also receive finite temperature corrections. Those are captured to a very good approximation by renormalizing all coupling constants at the scale of the first Matsubara mode, $\mu_R=2\pi T$, using renormalization group equations in vacuum~\cite{Giudice:2003jh}. \\

\begin{figure}[t!]
\begin{center}
\includegraphics[width=0.9\columnwidth]{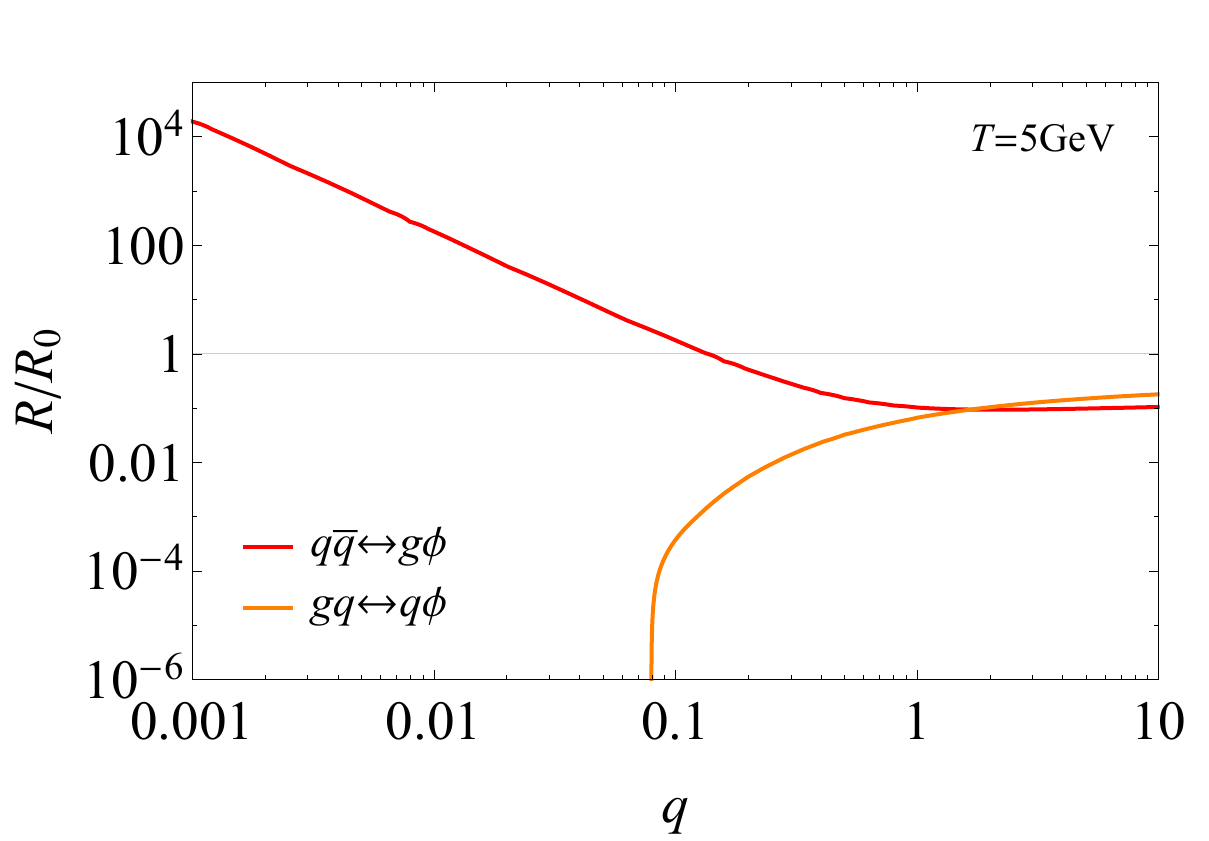}
\caption{Finite-temperature corrected QCD rate for $\phi$ production ($R$) relative to their  counterparts in vacuum ($R_0$).} 
\label{fig:ratesratio}
\end{center}
\end{figure}

There are several important implications of the plasma effects described above on mediator production. First of all, the quark thermal mass of $\mathcal{O}(g_sT)$ regulates the forward-enhancement of $t$-channel diagrams and thus strongly suppresses the production cross section compared to the zero-temperature limit. This is particularly visible at large $q\gtrsim 1$.
 In the opposite limit of small $q\lesssim 0.1-1$,  thermal masses affect the production rates in different and much more dramatic ways. For the $\phi q\to gq$ process, the large thermal mass of the gluon in the final state requires a highly energetic initial quark which is Boltzmann suppressed, causing the exponential drop below $q\sim 0.1$. The $\phi g \to q\bar q $ process however shows a strong enhancement  relative to the zero-temperature limit below $q\sim0.1$. This is more conveniently understood considering the direct process $q\bar q\to \phi g$. 
If the light $\phi$ particle emitted from the intial quark states is sufficiently soft, it becomes possible, since $m_g>2m_q$ within the plasma, that the gluon produced from the annihilating $q\bar q$ pair go on-shell, which strongly enhances the $2\to 2$ scattering amplitude. These effects of thermal masses are illustrated in Fig.~\ref{fig:ratesratio} which shows the ratio of the reaction rates in Eq.~\ref{coll2to2}, calculated with and without thermal masses.\footnote{In the absence of thermal masses, we regulate the forward-singularity of $t$-channel diagrams by cutting off phase space regions where the transfered momentum is less than $\Lambda_{\rm QCD}$.}
 
Second, since the gluon thermal mass is always larger than twice that of the quark, opening a new production channel from the decay $g\to q\bar q \phi$ which is absent at zero-temperature. Note that the photon thermal mass, which emerges from QED  interactions with charged particles in the plasma, is always too small to allow for the decay $\gamma\to q\bar q\phi$.

The impact of plasma effects on the $\phi$ distribution resulting from Eq.~\eqref{Beqphi} is illustrated on Fig.~\ref{fig:fphi} for $T=5\,{\rm GeV}$, where the solid black (orange) line represents the distribution with (without) including thermal corrections. For $2\to 2$ processes the inclusion of thermal masses suppresses by a factor of $\mathcal{O}(10)$ the production of $\phi$ particles with momentum larger than temperature. At small momenta, however, thermal masses strongly enhances $\phi$ production due to significant emission of soft $\phi$ particles together with gluons. This enhancement in the production rate allows $\phi$ to reach equilibrium at small momenta much faster than in the absence of thermal corrections since, in this case, the production peaks at momenta of $\sim T$. Consequently, the small momentum region is less populated, as shown with the orange line in Fig.~\ref{fphi}.      The gluon decay contribution is typically much smaller than the scattering contributions and quickly becomes inefficient in producing energetic $\phi$ particles due to phase space limitation.  The latter contribution drops from about 30\% at $p\sim T$ to roughly 10\% at high momenta. \\

\subsection{Simplifying assumptions}

The full resolution of the Boltzmann equations is rather cumbersome. However a relatively accurate result for the relic density can be obtained upon making several simplifying assumptions.

\begin{figure}[h]
\begin{center}
\includegraphics[width=0.9\columnwidth]{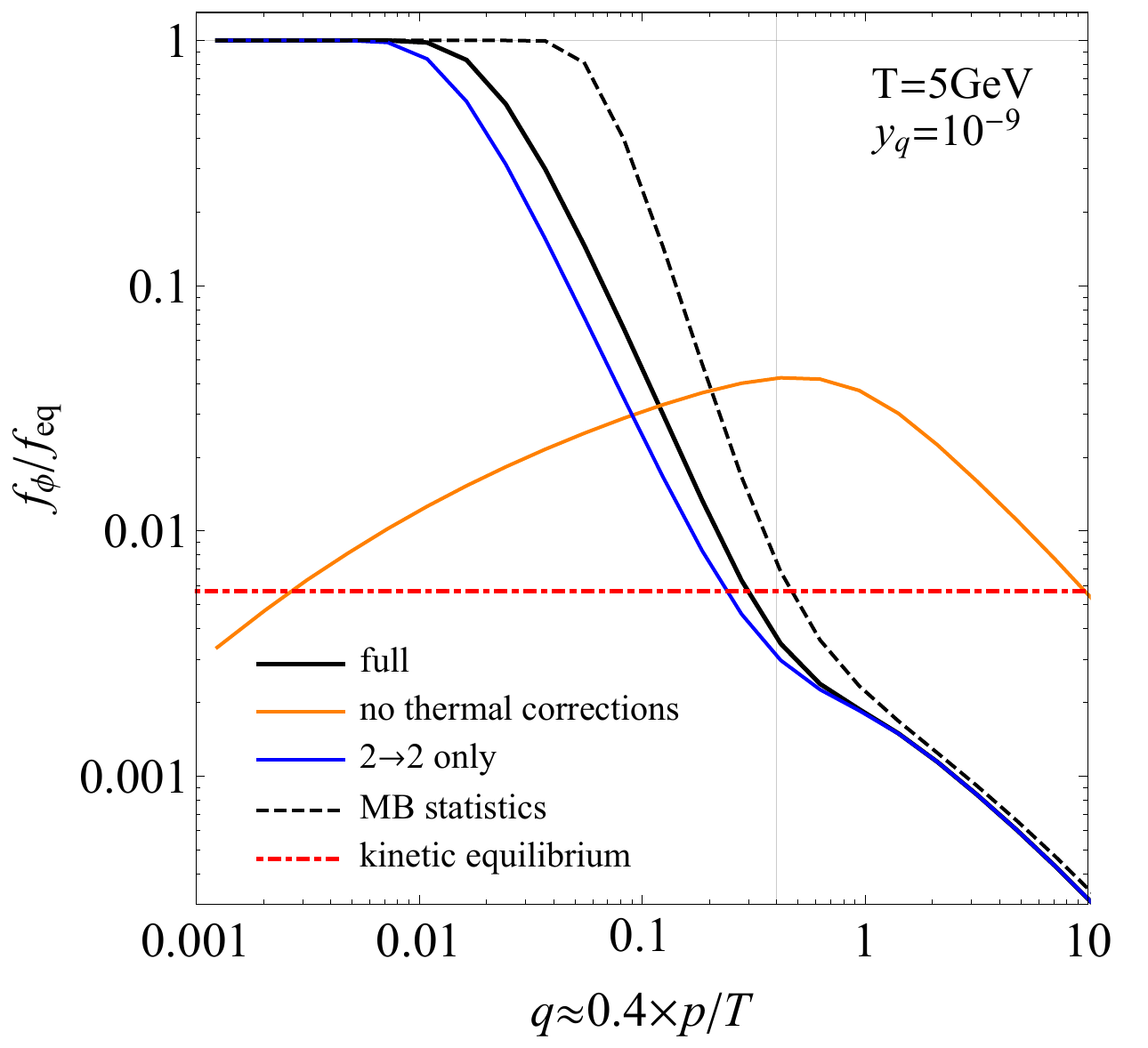}\\
\caption{
Momentum distribution of the mediator normalized to the equilibrium distribution at $T=5\,$GeV, assuming $y_q=10^{-9}$. The exact treatment which includes finite temperature corrections, quantum statistical distributions and thermal gluon decays (black) is compared  with the various approximations : neglecting thermal gluon decay (blue) or finite temperature corrections altogether (orange), assuming Maxwell-Boltzmann statistical distributions for all particles (dashed black) and in the kinetic equilibrium approximation (dot-dashed red).The vertical gray line denote the typical momentum required for DM production $p\simeq m_\chi=T$.}  \label{fig:fphi}
\end{center}
\end{figure}

First of all, we evaluate the impact of the choice of statistical distributions to describe particles in the plasma. The convenient assumption that particles follow a Maxwell-Boltzmann (MB) distribution is typically not justified for the freeze-in mechanism where most  DM particles are produced from collisions of very relativistic particles. As shown in Fig.~\ref{fig:fphi}, making the approximation that all particles have MB distributions would overestimate the production of $\phi$ particles by more than a factor 2  for $p\approx T$ and by about 10\% for much larger momenta. Note that using the MB distribution does  not have a strong impact on the gluon decay contribution~\cite{Belanger:2018mqt}.

Second, a simple approximation would be to assume that $\phi$ is in kinetic equilibrium with the thermal bath~\cite{Hambye:2019dwd}. In this  case $f_\phi/f_{\rm eq}$ is independent of momentum and simply given by the ratio $n_\phi/n_{\rm eq}$ of the $\phi$ number density, thus avoiding having to solve the unintegrated Boltzmann equation in Eq.~\eqref{Beqphi2}. This approximation is not justifed {\it a priori}, unless  $n_\phi\approx n_{\rm eq}$, because there is no elastic scattering rate between $\phi$ and SM particles that is faster than the Hubble rate. Moreover, since $\phi$ production rates are faster at low momentum, the kinetic equilibrium approximation largely overestimates (underestimates)  $f_\phi$ at high (low) momenta by several orders of magnitude, as shown in Fig.~\ref{fig:fphi} with the horizontal red dot-dashed line. 
As argued in Appendix ~\ref{app:approx} the peak of DM production through fusion of out-of-equilibrium $\phi$ pairs occurs for one $\phi$ particle with a large momentum of $\mathcal{O}(m_\chi)$ colliding with another one  nearly at rest. Therefore, within the kinetic equilibrium assumption, there is a large compensation between the $\phi$ distribution at small and large $q$. As a result of this partial cancellation the kinetic approximation allows to estimate the DM relic density from out-of-equilibrium mediator fusion within an $\mathcal{O}(1)$ factor (see below).

\section{DM production}
\label{sec:freezein}
We assume negligible initial abundance for the dark sector at the end of inflation, $n_\chi=n_\phi=0$ at $T=T_R$ with $T_R$ denoting the reheating temperature. In constrast with thermal production, the DM relic is produced by the so-called freeze-in mechanism~\cite{McDonald:2001vt,Hall:2009bx} through feeble interactions with the thermal bath (during the radiation-dominated era). There are two possible channels for DM production: $q\bar q\to \chi\bar\chi$ (with $\phi$ in the $s-$channel) and $\phi\phi\to \chi\bar\chi$ (with $\phi$ in the $t,u-$ channels) where the $\phi$ density is produced from thermal collisions of SM fields (see below).

The DM yield $Y_\chi\equiv n_\chi/s$ where $n_\chi$ is the DM number density and $s$ is the entropy density associated with the SM degrees of freedom, is governed by the following Boltzmann equation,
\beq\label{BEq}
s\bar H x \frac{dY_\chi}{dx} = \gamma_{q\bar q\leftrightarrow\chi\bar\chi}+\gamma_{\phi\phi\leftrightarrow\chi\bar\chi}\,,
\eeq
where $x\equiv m_\chi/T$, $\bar H\equiv H/[1-(d\log h_{\rm eff}/d\log x)/3]$, $H$ being the Hubble parameter,
and $\gamma_{A\leftrightarrow B} \equiv \gamma_{A\to B}-\gamma_{B\to A}$. The $\gamma$'s are the so-called (integrated) collision terms associated with the production processes described above ($A=q\bar q$ or $\phi\phi$) and their depletion counterparts.

The total DM energy density today is obtained from integrating Eq.~\eqref{BEq} between $T_R$ and $T_0\sim2.7\,$K (the photon temperature today)
\beq
\Omega_\chi \simeq \frac{m_\chi s_0}{\rho_c}\int_{x_R}^{x_0} dx\, \frac{dY_\chi}{dx}\,,
\eeq
where $s_0\approx2.89\times 10^9/$m$^{-3}$ and $\rho_c\approx10.54 h^2\,$GeV$/$m$^{-3}$ are today's  entropy and critical energy densities of the universe, respectively. $h\approx0.674(5)$~\cite{Planck2018} is related to the value of the  Hubble parameter today as $H_0=100h\,$km/sec/Mpc. 

The value of $T_R$ is somewhat arbitrary. The simplified model under consideration being only valid below the EW scale we set $T_R=100\,$GeV for consistency. Higher values of $T_R$ would require to embed the interaction Lagrangian in Eq.~\eqref{Lint} into a specific UV complete theory respecting the  SU(2)$_L\times$U(1)$_Y$ invariance of the SM. Note that for $m_\chi\ll T_R$, DM is dominantly produced at much lower temperatures $T\sim m_\chi$ where the relevant dynamics is well described by Eq.~\eqref{Lint} and the precise value of $T_R$ irrelevant. However, production of heavier DM particles would be strongly suppressed. Nevertheless, the freeze-in mechanism for $m_\chi\gtrsim 50\,$GeV is well covered by direct detection~\cite{Hambye:2018dpi} and most probably excluded by Xenon1T~\cite{Aprile:2018dbl}, see also section~\ref{sec:detection}.\\

We solve Eq.~\eqref{BEq} neglecting reverse processes where DM annihilates back into $q\bar q$ and $\phi\phi$. This is certainly a justified approximation for the hadronic channel since $\gamma_{\chi\bar\chi\to q\bar q}/\gamma_{q\bar q\to\chi\bar\chi} \sim\mathcal{O}(n_\chi^2/n_{\rm eq}^2)$ and $n_\chi\ll n_{\rm eq}$ at all times. The situation is less clear for mediator channel though, in particular because, as we show below, DM could be efficiently produced  also in the case that $\phi$ is not in equilibrium with the thermal bath and $n_\phi\ll n_{\rm eq}$. We verified numerically that the number density $\bar{n}_\phi$ of $\phi$ particles with energy above $m_\chi$ is larger than $n_\chi$ by a factor of $\mathcal{O}(10^3)$ or more in regions of parameter space where $\phi\phi\to \chi\bar\chi$ is dominant. Hence  $\gamma_{\chi\bar\chi\to \phi\phi}/\gamma_{\phi\phi\to\chi\bar\chi}\sim \mathcal{O}(n_\chi^2/\bar{n}_\phi^2)\ll 1$ whenever relevant and the reverse process is also negligible in this case.\\

\begin{figure}[h]
\begin{center}
\includegraphics[width=0.9\columnwidth]{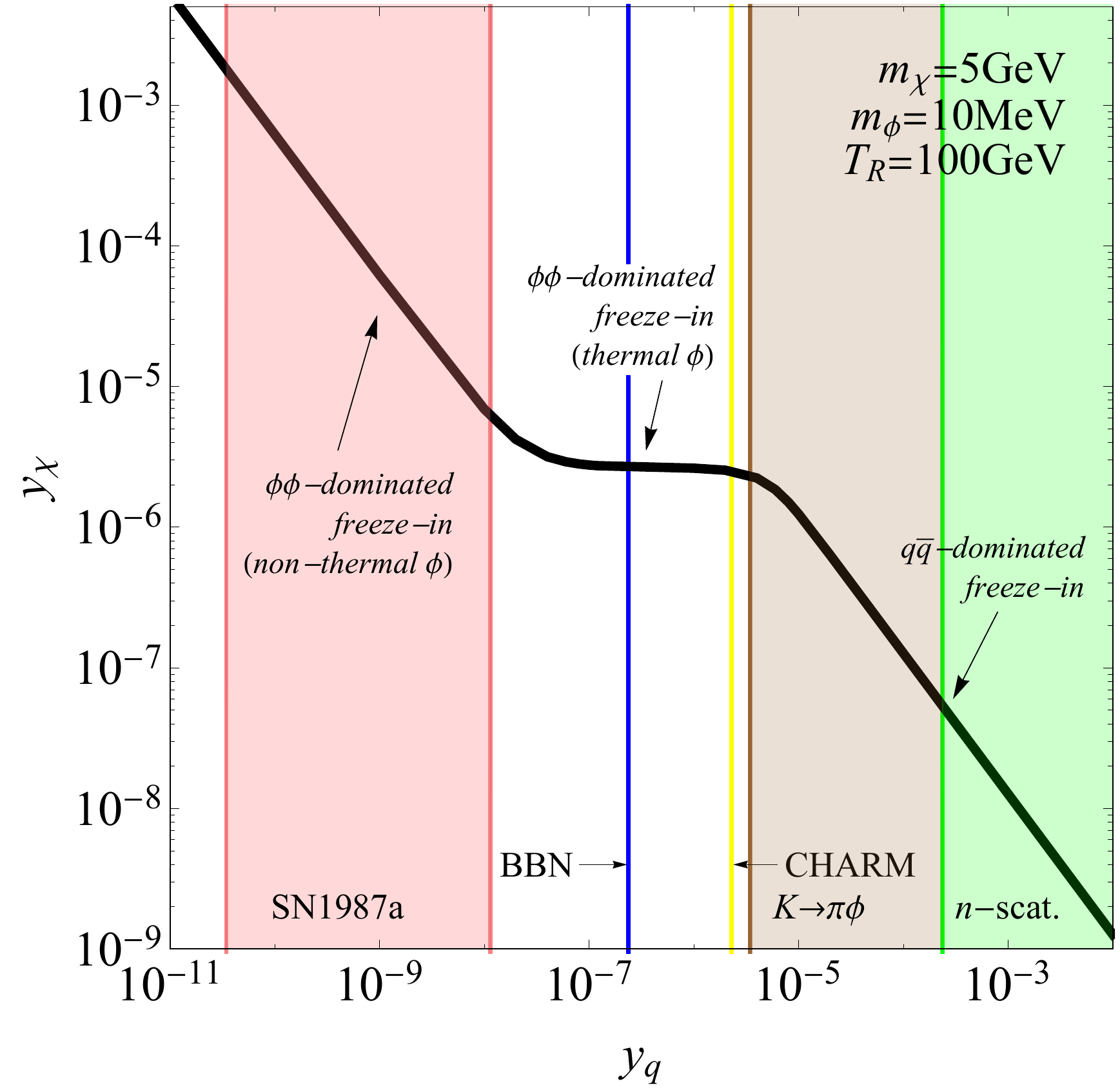}
\caption{Phase diagram of the relic density for a $5\,$GeV mass DM and a MeV-scale mediator. The  contour in black reproduces $\Omega_{\rm DM}h^2=0.118$. Below (above) this line, DM is under (over) abundant. Here $y_q$ is defined at the scale $m_\chi$. The vertical bands correspond to the constraints described in Section~\ref{sec:constraints}, see text for details.}
\label{fig:phasediag}
\end{center}
\end{figure}

For illustration, we show in Fig.~\ref{fig:phasediag} the ``phase diagram" in the $y_q-y_\chi$ coupling plane resulting from the calculation of the relic density of $5\,$GeV mass DM in the model described in the previous section. One distinguishes three different regimes for DM production depending on the value of the quark-mediator coupling. 

For relatively large values of $y_q$ the dominant DM production mechanism is directly from collisions of thermal SM particles, through $q\bar q\to \chi\bar\chi$ whose cross section (as well as $\Omega_\chi$) scales like $(y_qy_\chi)^2$.

As the quark coupling is decreased, SM collisions are less and less frequently producing DM particles and below a critical value of $y_q^{\rm crit}$ (the precise value slightly varies with $m_\chi$) collisions of mediator particles become the dominant production mechanism. Since $\phi$ is assumed with negligible initial density, the efficiency of this process is determined by how much $\phi$ particles are produced from SM collisions. 
For values not too far below $y_q^{\rm crit}$, the quark coupling is typically still sufficiently large so that the mediator reaches equilibrium with the thermal bath before DM production effectively starts at $T\sim m_\chi$. In that case, the density of $\phi$ no longer depends on $y_q$ and $\Omega_\chi$ scales like $y_\chi^4$. Hence the plateau in the phase diagram of Fig.~\ref{fig:phasediag}. 

For even smaller values of $y_q\lesssim y_q^{\rm eq}$, where $y_q^{\rm eq}$ is the minimal coupling value needed to keep $\phi$ in thermal equilibrium with SM, the $\phi$ production rate is too slow, such that the mediator is out-of-equilibrium during DM production. In this case, the $\phi\phi\to\chi\bar\chi$ rate is suppressed by the (square of the) small density of non-thermal $\phi$. Nevertheless, this mechanism still dominates over direct production from SM collisions. This is understood as follows. Far from equilibrium, {\it ie.} for $y_q\ll y_q^{\rm eq}$, the momentum distribution of $\phi$ is proportional to the production rate and scales as $f_\phi\sim (y_q/y_q^{\rm eq})^2f_{\rm eq}$ since the rate is dominated by single production processes. As a result, the $\phi\phi\to \chi\bar\chi$ contribution to the relic density scales as $(y_qy_\chi/y_q^{\rm eq})^4$ in this regime, and the ratio of collision terms in Eq.~\eqref{BEq} is (roughly) $\gamma_{\phi\phi\to\chi\bar \chi}/\gamma_{q\bar q\to \chi\bar\chi}\sim (y_qy_\chi)^2/(y_q^{\rm eq})^4$. The $q\bar q$-dominated freeze-in mechanism requires $y_qy_\chi\sim 10^{-11}$ in order to reproduce the observed DM relic density, while typically $y_q^{\rm eq}\approx  10^{-7}$. Hence, the freeze-in production of DM from collisions of non-thermal $\phi$ particles dominates over the direct contribution from SM collisions by a factor of $\sim10^5$. 

To conclude this section, we stress the importance of using the full solution of the Boltzmann equation for $\phi$ in computing the DM relic density. Detailed comparison reveals that assuming a kinetic equilibrium distribution for the mediators  overestimates $\Omega_\chi$ by a factor $\sim2$ for $y_q < 10^{-9}$ while the discrepancy with the full calculation rapidly disappears  for coupling values large enough so that $\phi$ approaches thermal equilibirum.
As noted in Section~\ref{sec:phireezein}, using Maxwell-Boltzmann distributions leads to an overproduction of  $\phi$ particles. For $m_\chi=5\,$GeV, this overestimates the relic density by about 50\%  in a regime where $\phi$ is out-of-equilibirum. The neglect of finite temperature corrections and plasma effects also leads to an overproduction of $\phi$ particles that yields an ${\cal O}(1)$ increase in the relic density. On the other hand, if $\phi$ particles are in thermal equilibrium, the relic density is roughly 40\% higher when using the Bose-Einstein statistical distribution. Finally, when DM is mainly produced from u-quarks, its relic density increases more mildly, around 25\%, when using a MB distribution.

\section{Predictions for direct detection experiments}
\label{sec:detection}

A relic of DM particles of mass above the GeV-scale can be directly detected by observing scattering events on heavy nuclei~\cite{WittenDM}. For a scalar mediator the expected signal is spin-independent (SI) with differential rate $dR/dE_R$ as a function of the nuclear recoil energy $E_R$ given by~\cite{Lewin:298578},
\beq\label{dRdER}
\frac{dR}{dE_R}=\frac{\rho_0\bar\sigma_{\rm SI}N_A}{\sqrt{\pi}v_0m_\chi \mu_{\chi N}^2}F^2(q)\eta(q^2)\times \frac{m_\phi^4}{(q^2+m_\phi^2)^2}\,,
\eeq
where $q^2\equiv \sqrt{2m_NE_R}$ is the momentum transfered, $\rho_0=0.3 {\rm GeV}/{\rm cm}^3$ is the DM energy density today, $m_N$ is the mass of the target nucleus, $\mu_{\chi N}\equiv m_\chi m_N/(m_\chi+m_N)$ is the reduced mass of the DM-nucleus system and $N_A$ is the Avogadro constant.  
$F(q)$ is a nuclear form factor that describes the loss of coherence among nucleons at finite momentum transfer,  
while $\eta(q^2)$ captures the dependence on the DM velocity distribution. Their explicit functions are given in Appendix~\ref{app:dd}.

$\bar\sigma_{\rm SI}$ is the SI DM-nucleus scattering cross section evaluated at zero momentum transfer. In the limit of isospin symmetry it is related to the cross section on a single nucleon, say proton, as $\bar\sigma_{\rm SI}/\mu_{\chi N}^2=A^2\bar\sigma_{\rm SI}^p/\mu_{\chi p}^2$ where $A$ is the total number of nucleons in the target and $\mu_{\chi p}=m_\chi m_p/(m_\chi+m_p)$ is the reduced mass of DM and the proton. In our simplified model, assuming $m_\phi\ll m_\chi$, we have 
\beq
\bar\sigma_{\rm SI}^p= \frac{y_p^2y_\chi^2\mu_{\chi p}^2}{\pi m_\phi^4}\,, 
\label{eq:sig_si}
\eeq
where $y_p$ is defined in Eq.~\ref{eq:yp}. 
Finally, the last term on the right-hand side of Eq.~\eqref{dRdER} parameterizes the $t$-channel propagator of the mediator. For $q^2\ll m_\phi^2$, the DM-nucleus scattering is well described by a contact interaction, which is the implicit assumption behind the limit on $\bar\sigma_{\rm SI}^p$ (as a function of $m_\chi$) presented by all DD experiments. However for $m_\phi\lesssim q_{\rm max}\sim \mathcal{O}($GeV), the $q^2$-dependence of the cross section is not negligible, and limits assuming contact interactions no longer apply. Nonetheless, the DD sensitivity for light mediators can be estimated by recasting existing limits based on the event rate expected from DM scattering in a given experiment. We have followed the recasting procedure of micrOMEGAs~\cite{Belanger:2020gnr} for Xenon1T~\cite{Aprile:2018dbl} and DarkSide50~\cite{Agnes:2018ves}.
In the low mass region, the latter is superseded by two analyses from Xenon1T using the S2 signal only~\cite{Aprile:2019xxb} and taking advantage of the Migdal effect~\cite{Aprile:2019jmx}. To estimate the projected sensitivity of SuperCDMS~\cite{Agnese:2016cpb}, the expected event rate is computed using
\beq
R=\int dE_R\, \epsilon(E_R) \frac{dR}{dE_R}\,
\eeq
where $\epsilon(E_R)$ denotes  the detection efficiency. We assume that the efficiency vanishes below the  nuclear energy threshold of 0.04keV  and increases linearly to reach 
 85\% at  2keV, for higher energies we take a constant efficiency~\cite{Agnese:2016cpb}.  Note that the exact shape of the efficiency curve at low nuclear recoil energies strongly affects the event rate, since the energy distribution for a light mediator peaks at low energies.

\begin{figure}[h]
\begin{center}
\includegraphics[width=0.9\columnwidth]{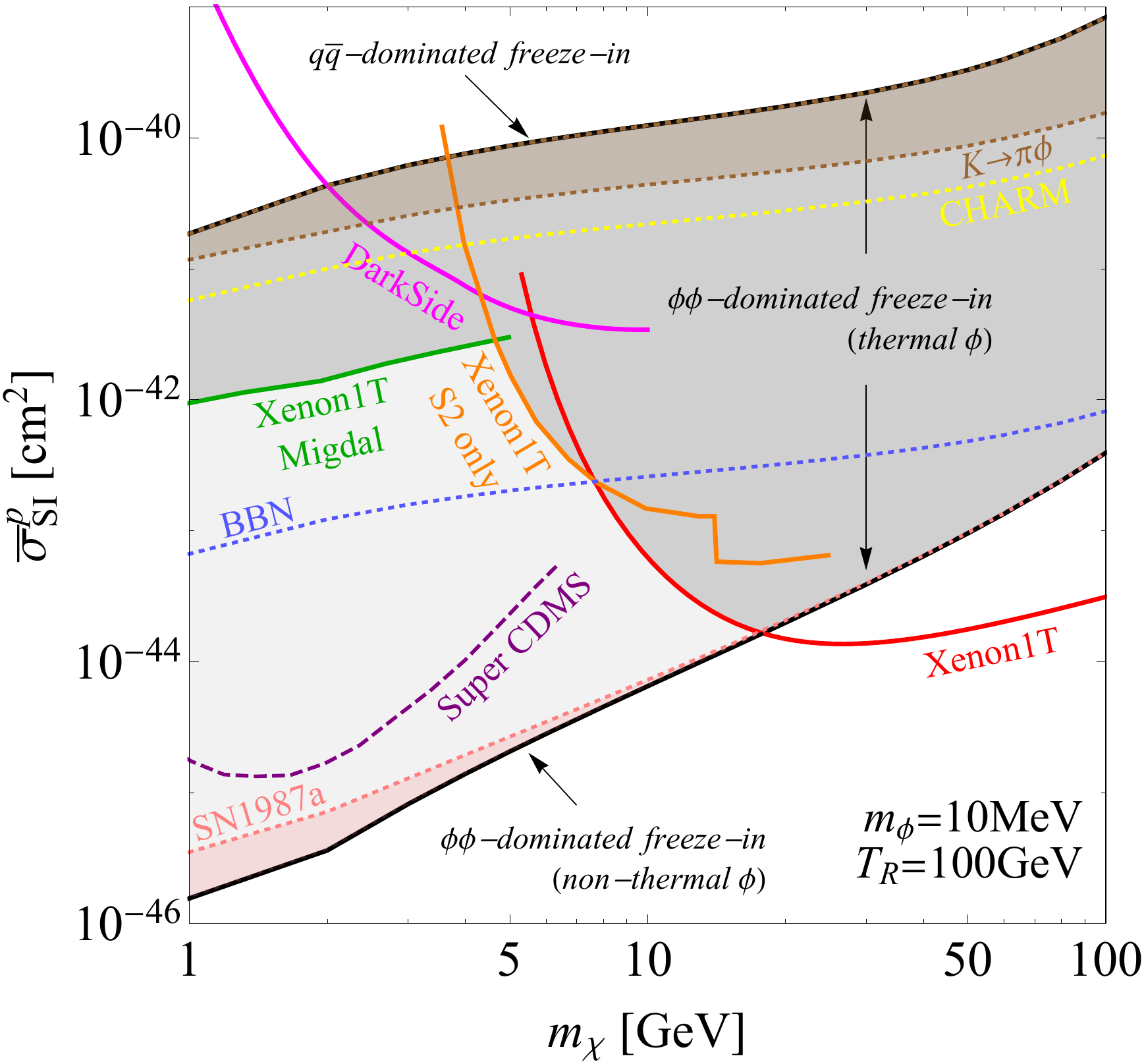}
\caption{Freeze-in prediction for the cross section of DM spin-independent scattering of proton at zero-momentum transfer. Experimental limits and future sensitivities have been recast to account for the MeV-scale mediator. See text for details.}
\label{fig:dd}
\end{center}
\end{figure}

The freeze-in prediction for the SI cross-section strongly depends on whether the relic abundance is dominated by $q\bar q$-initiated or $\phi\phi$-initiated collisions in the early universe. In the first case, both $\Omega_\chi$ and $\bar \sigma_{\rm SI}^p$ depend on the same combination of couplings, $(y_q y_\chi)^2$, such that the relic density uniquely determines the direct detection signal for a fixed DM mass. This prediction is represented by the upper black line on Fig.~\ref{fig:dd}. In the second case, when mediator collisions dominate  the freeze-in production of DM, the product of couplings $y_qy_\chi$ could be much smaller, and its value depends on whether the mediator is in thermal equilibrium with the SM or not, see Fig.~\ref{fig:phasediag}. When freeze-in is dominated by non-thermal $\phi$, $\Omega_\chi\propto (y_qy_\chi)^4$ and the product $y_qy_\chi$ is also fixed. Thus, the relic density also makes a unique prediction for $\bar \sigma_{\rm SI}^p$ in this case, which is represented by the lower black line on Fig.~\ref{fig:dd}. Conversely, when the mediator is in thermal equilibrium during DM production, $\Omega\propto y_\chi^4$ while the quark coupling is in the range $y_q^{\rm eq}\lesssim y_q\lesssim y_q^{\rm crit}$. Hence, the direct detection cross section predicted by the relic density is not unique, but rather lies within the entire interval between the predicted value of $q\bar q$-dominated regime (above) and that of the $\phi\phi$-dominated one with non-thermal $\phi$ (below). This results in the gray-shaded band shown in Fig.~\ref{fig:dd} which approximately spans three to five orders of magnitude, depending on the  DM mass. Note that the $\bar \sigma_{\rm SI}^p$ range predicted by freeze-in is narrower for larger values of $m_\chi$, which is simply due to the fact that the mediator requires a larger coupling to the SM in order to reach equilibrium before DM production starts.

The predicted direct detection signals are already well covered by existing experiments. In particular, the current limits from Xenon1T rule out the parameter space consistent with freeze-in for DM masses above $30\,$GeV and partly covers the $\phi\phi$-dominated freeze-in down to its threshold sensitivity, corresponding to $m_\chi\approx6\,$GeV. For lower DM masses, the new analyses by Xenon1T based on the Migdal effect or exploiting the S2 signal only exclude the $q\bar q$-dominated freeze-in as well as part of the parameter space of the $\phi\phi$-dominated regime. This region is also excluded partly by DarkSide50. Moreover, future experiments will significantly improve the coverage for light DM. For instance, the projected reach for SuperCDMS will  allow to probe a significant fraction of the freeze-in prediction below $5\,$GeV. 

Finally, several terrestrial, astrophysics and cosmology constraints, which we summarise in the next section for completeness, can be imposed on a MeV-scale hadrophilic scalar. Imposing all these constraints at face value  for $m_\phi= 10\,$MeV severely restricts the range for the quark-mediator coupling, such that most of the mediator-dominated regime would be excluded. The narrower band of the SI cross-sections that reproduce the relic abundance and satisfy these constraints  corresponds to the upper part of the light grey area above the BBN line in Fig.~\ref{fig:dd}. This region is significantly enlarged to the whole light grey area when
 the BBN bound is evaded in the presence of an additional decay channel for the mediator into neutrinos.

\section{Other constraints}
\label{sec:constraints}
Other constraints exist on individual couplings of a light mediator to SM quarks, both from terrestrial experiments and astrophysics and cosmological observations. See Refs.~\cite{Zurek-lightDM,Batell:2018fqo} for detailed reviews. For completeness we quickly decribe below the constraints shown in Figs.~\ref{fig:phasediag},\ref{fig:dd} that are relevant to our scenario.\\

Several laboratory experiments are sensitive to light particles coupled to quarks. Neutron-scattering experiments at low energies are sensitive to the coupling of $\phi$ to neutrons. The strongest constraint for a 10$\,$MeV mass~\cite{Frugiuele:2016rii} arises from analyzing the momentum distribution of keV-scale neutrons scattered off lead nuclei, giving $y_n\lesssim 1.5\times 10^{-3}$~\cite{PhysRevLett.68.1472}. This bound translates to $y_q(2\,$GeV$)\lesssim 2.9\times 10^{-4}$ for a scalar coupled to $u$-quark only~\cite{Shifman:1978zn,Bhattacharya:2016zcn}. 

Additional constraints come from rare meson decays.
 Light mediators coupled to quarks can be produced on-shell in $B\to K\phi$ and $K\to \pi \phi$ decays~\cite{Willey:1982mc}. For the coupling values we envisage here $\phi$ is stable on collider scales and would appear as missing energy in the decays. Due to the heavy bottom mass, the $B\to K$ transition is induced at one-loop by an electroweak penguin diagram. This penguin is suppressed by the light $u$-quark mass and small CKM matrix elements $V_{us}V_{ub}$, which makes it negligible given the current experimental bound on such decay~\cite{Lees:2013kla}. However, the $K\to \pi$ transition receives a less suppressed tree-level contribution from the chiral Lagrangian~\cite{Batell:2018fqo}. The limit on the BR($K^+\to \pi^+\nu\bar\nu$)~\cite{Kpinunu} provides a strong constraint from meson decays, yielding $y_q(2\,$GeV$)\lesssim 4.2\times 10^{-6}$ for $u$-quark. The forthcoming NA62 experiment is expected to improve the sensitivity in the $K^+\to \pi^+\nu\bar\nu$ channel by a factor of $\sim3$~\cite{NA62Kpinunu}.
Light mediators coupled to quarks are also constrained from proton beam dump experiments. In particular, the axion-like particle search at the 400$\,$GeV SPS by the CHARM collaboration~\cite{Bergsma:1985qz} can be used to constrain the process  $\eta \rightarrow \pi \phi$ where $\phi$ decay into two photons~\cite{Batell:2018fqo}. This search yields the strongest  upper limit from meson decay,   $y_q(2\,$GeV$)\lesssim 2.8\times 10^{-6}$ for $u$-quark. Note however that this constraint can be relaxed if  $\phi$ is allowed to decay into an invisible channel, like neutrinos.

Light bosons coupled to nucleons can be emitted in stars. Below a critical coupling value, the emitted bosons interact so weakly with the stellar medium that they escape the star without being reabsorbed, thus contributing to its cooling. Lack of evidence of such additional energy loss mechanisms in several stellar systems thus constrains the coupling of light bosons~\cite{Raffelt:1996wa}. For large enough couplings, the new bosons are efficiently reabsorbed and trapped in the stellar medium, thus no longer contributing to energy losses. Horizontal branch and red giant stars are too cold to emit 10$\,$MeV-scale bosons. Those can however be constrained from supernova 1987A (SN1987A) whose temperature reached $T\sim 30\,$MeV, excluding $u$-quark coupling values  in the range $4.2\times 10^{-11}\lesssim y_q(2\,$GeV)$ \lesssim 1.4\times 10^{-8}$.\\

Light bosons with tiny coupling to SM fields typically live long enough to leave traces in well-understood late cosmological phenomena, such as BBN or the cosmic microwave background (CMB). If the mediator survives until BBN, its $\phi\to \gamma\gamma$ decay could inject electromagnetic energy in the thermal bath, hence increasing its entropy density and (if the decay products are sufficiently energetic)  dissociating the freshly formed light elements. Given its very small scattering cross-section with the SM, the mediator will decouple relativistically from the thermal bath at $T\sim m_\phi=10$\,MeV. Then, in order to avoid strong alteration of the standard BBN predictions for the abundances of light elements, its relic must decay away before the onset of the first nuclear reactions at $t\simeq1\,$sec. From Eq.~\eqref{phiwidth}, this implies $y_q(2\,$GeV$)\gtrsim 2.9\times 10^{-7}$. 
Note that this constraint can be evaded by shortening the lifetime of the mediator through either increasing its mass or opening an additionnal decay channel into neutrinos. In the latter case, we checked,  using the {\tt alterBBN}~\cite{Arbey:2018zfh} code, that the neutrinos produced from the decay of a 10$\,$MeV mediator thermalize before neutrino decoupling and do not spoil BBN predictions.

\section{Discussion}

In the above analysis, we concentrated on the specific case of a $10\,$ MeV scalar mediator that couples to DM and $u$-quarks. We also implicitly assumed a coupling to SM neutrinos whenever necessary to avoid cosmological constraints. However, the mechanism of DM freeze-in production from fusion of out-of-equilibrium mediators is more generic. First of all, our results for the DM relic density would equally apply in cases where $\phi$ couples to any of the light quark flavors. Moreover,  DM phenomenology remain valid as long as the mediator mass is below $\sim100\,$MeV. Indeed such a light mediator has little impact on the relic density, since $m_\chi\gg m_\phi$, as well as on  predictions and limits from direct detection, since the scaling of the cross section $\bar\sigma^p_{SI}\propto m_\phi^{-4}$ holds in this mass range. The mass of the mediator and its possibility to decay into a neutrino channel do, however, affect other constraints, most notably those from BBN. In the absence of the neutrino channel, a mediator's lifetime $\tau_\phi \lesssim 1$ sec can be achieved with a large enough coupling $y_q$, more precisely for $y_q(m_\phi/10\,$MeV$)^{3/2}>2.9 \times 10^{-7}$.  While  BBN constraints restrict the out-of-equilibrium regime for lighter mediators,  $m_\phi> 30$ MeV allows for sufficiently smaller values of $y_q$ such that the out-of-equilibrium regime opens up significantly. This also 
implies a wider region in direct detection of light dark matter, with  $m_\chi\lesssim 6\,$GeV, that is free of constraints. Note also that for $m_\phi> 30$ MeV,  the supernova constraint disappears thus further relaxing constraints on the whole out-of-equilibrium window.

In summary,  we performed a detailed calculation of DM production via the production of out-of-equilibrium mediators by solving the unintegrated Boltzmann equation for the latter, instead of making the kinetic equilibrium approximation, and including the effect of thermal masses and of quantum statistics.  Each of these effect has a large impact on the prediction of the dark matter relic density. We also showed using a simplified model that while  this mechanism faces cosmological constraint it can be probed by direct detection experiments. Increasing the sensitivity of direct detection experiments at low masses is however crucial to completely cover the pure  freeze-in via out of equilibrium region in the future.

\begin{acknowledgments}
We thank T. Hambye and M. Tytgat for fruitful discussions and A. Arbey for providing useful informations on {\tt alterBBN}.
This work was funded by RFBR and CNRS, project number 20-52-15005 and by a grant AAP-USMB. 
GB and AP  would like to thank the  Paris-Saclay Particle Symposium 2019 with the support of the P2I and SPU research departments and of
the P2IO Laboratory of Excellence (program "Investissements d'avenir"
ANR-11-IDEX-0003-01 Paris-Saclay and ANR-10-LABX-0038) for their hospitality and support during the completion of this work.
BZ is supported by the Programa Atracci\'on de Talento de la Comunidad de Madrid under grant n. 2017-T2/TIC-5455, from the Comunidad de Madrid/UAM “Proyecto de J\'ovenes Investigadores” grant n. SI1/PJI/2019-00294, from Spanish “Proyectos de I+D de Generacion de Conocimiento” via grants PGC2018-096646-A-I00 and PGC2018-095161-B-I00.

\end{acknowledgments}

\appendix
\section{Reaction rate integrals for mediator production}
\label{app:collisionterms}

We elaborate on the calculation of the reaction rates relevant for mediator production in the early universe.

\subsection{Scattering}

The rate of the $2\to 2$ scattering reaction $12\leftrightarrow 3\phi$ is
\beq\label{coll2to2App}
R_{\rm scat} = \frac{g_3}{(2\pi)^3} \int d^3p_3 f_3\sigma_{3\phi\to 12}v_{\rm M\o l}\,,
\eeq
see Eq.~\eqref{coll2to2}. The  M\o ller velocity $v_{\rm M\o l}=F/(EE_3)$ where $F=\sqrt{s}\, p_*(s)$ is the (Lorentz-invariant) flux of incoming particle, expressed in terms of the total energy $\sqrt{s}$ and the $\phi$ momentum $p_*(s)=\sqrt{s}/2[1-(m_3+m_\phi)^2/s]^{1/2}[1-(m_3-m_\phi)^2/s]^{1/2}$ in the center-of-mass frame, while $E$ and $E_3$ are the energies of $\phi$ and particle 3 in the plasma frame. The differential element in Eq.~\eqref{coll2to2App} writes $d^3p_3=2\pi p_3^2dp_3d\cos\theta$ where $\cos\theta$ can be identified to the angle between $p_3$ and $p$ in the plasma frame. Using $s=m_3^2+m_\phi^2+2(EE_3-p p_3\cos\theta)$, the $\cos\theta$ integral can be traded for an integral over $s$, yielding
\beq\label{coll2to2App2}
R_{\rm scat} = \frac{g_3}{8\pi^2Ep}\int ds \sqrt{s}\,p_*\,\sigma_{3\phi\to 12} \int dE_3 f_3\,,
\eeq
where $s\geq {\rm max}[(m_1+m_2)^2,(m_3+m_\phi)^2]$ 
. The boundary values $E_3^{\rm min,max}=E_3^{\rm min,max}(s,E)$ are obtained as follows. For fixed $s$ and $E$, $E_3$ reaches its extremal values when particle 3 and $\phi$ are collinear. In this case, $E_3$ relate to $s$ and $E$ through a Lorentz boost transformation,
\beq
E_3= E_{3*}\cosh(y)-p_*\sinh(y)\,,
\eeq
where $E_{3*}=(s+m_3^2-m_\phi^2)/2\sqrt{s}$ and $p_{3*}=-p_*$ are the energy and momentum of particle 3 in the center-of-mass frame and $y$ is the rapidity of the center-of-mass, which satisfies $E=E_*\cosh(y)+p_*\sinh(y)$ with $E_*=\sqrt{s}-E_{3*}$. There are two independent solutions, 
\beq
y_\pm = \log\left(\frac{E\pm p}{E_*+p_*}\right)\,,
\eeq
where $y_+$ ($y_-$) is reached when the momenta of particle 3 and $\phi$ are parallel (antiparallel), and the energy of particle 3 in the plasma frame is $E_3^{\rm max}$ ($E_3^{\rm min}$). 

\subsection{Decay}

The rate of the 3-body decay reaction $1\to 23\phi$ is 
\beq\label{coll1to3App}
R_{\rm decay}= g_1e^{E/T}\int d^3p_1 f_1 \frac{d\Gamma_{1\to23\phi}}{d^3p}\,,
\eeq
see Eq.~\eqref{coll1to3}, where $\Gamma_{1\to 23\phi}$ is the decay width of particle 1 in the plasma frame, $p$ is the $\phi$ momentum in that frame and 
\beq
\frac{d\Gamma_{1\to 23\phi}}{d^3p} = \frac{1}{2\pi E p}\frac{d\Gamma_{1\to 23\phi}}{dE}\,.
\eeq
In order to determine $d\Gamma_{1\to 23\phi}/dE$, we will first calculate the differential width in the rest frame of the decaying particle 1, $d\Gamma_{1\to 23\phi}^*/dE_*$, and boost it to plasma frame. First of all, the width is boosted by a Lorentz factor $d\Gamma_{1\to 23\phi}=m_1/E_1\times d\Gamma_{1\to 23\phi}^*$. Then, one needs to find how many $\phi$ particle with energy $E_*$ in the rest frame of particle 1 wind up with energy $E$ in the plasma frame.
The energies $E$ and $E_*$ are related by a boost transformation 
\beq\label{boostdecay}
E=E_*\cosh(y)-p_*\cos\theta_*\sinh(y)\,,
\eeq
with rapidity $y=1/2\log[(E_1+p_1)/(E_1-p_1)]$. $\theta_*$ is the angle between the $\phi$ momentum in the  particle 1 rest frame and the boost direction, given by particle 1 momentum.
Assuming the differential width is flat in $\cos\theta_*$, which is the case for the decay of a scalar or an unpolarized particle with non-zero spin, a $\delta$-distribution $d\Gamma/dE_*=A\delta(E_*-E_{0*})$ in the particle 1 rest frame yields a rectangular function in terms of $E$ in the plasma frame, $d\Gamma/dE=A/N[\Theta(E-E_{\rm min})-\Theta(E-E_{\rm max})]$ whose boundaries are defined by Eq.~\eqref{boostdecay} upon setting $\cos\theta_*=\pm 1$ and normalization is rescaled by the size of the rectangle $N=E_{\rm max}-E_{\rm min} = 2(E_{0*}^2-m_\phi^2)^{1/2}\sinh(y)$. Since any generic spectrum in $E_*$ can be decomposed as a (infinite) set of $\delta$-function peaking at different values of $E_*$, the boosted spectrum is simply given by adding up the rectangles, yielding
\beq
\frac{d\Gamma_{1\to 23\phi}}{dE}=\frac{m_1/E_1}{2\sinh(y)}\int \frac{dE_*}{\sqrt{E_*^2-m_\phi^2}}\frac{d\Gamma^*_{1\to 23\phi}}{dE_*}\,,
\eeq 
where the integration boundaries are $E\cosh(y)-p\sinh(y)\leq E_*\leq E\cosh(y)+p\sinh(y)$\,.\\

\section{Approximate $f_\phi$ solutions} 
\label{app:approx}

We derive analytical solutions for $f_\phi$ within some simple approximations. DM is dominantly produced at temperature  $T\sim m_\chi/3$, from collisions of very relativistic $\phi$ particles with momentum $p\gtrsim m_\chi\gg m_\phi$, which corresponds to $q\gtrsim 1$. 

Consider the reaction rate for $2\to 2$ processes of Eq.~\eqref{coll2to2App2} for large momentum $p\gg m_\phi,m_3\sim g_sT$. In that limit, $\exp(y_-)\simeq 0$ and $\exp(y_+)\simeq 2p/\sqrt{s}$ such that $E_3\gtrsim s/(4p)$. Assuming Maxwell-Boltzmann statistics for SM particles in the plasma, $f_3=\exp(-E_3/T)$ gives $\int dE_3 f_3\simeq T\exp[-s/(4pT)]$. For  high-energy scattering in the $t$-channel $\sigma_{3\phi\to 12}\sim c/s\log(s/m_0^2)$, where $c\sim g_s^2 y_q^2\ll 1$ and $m_0\sim g_s T$ is the mass of the exchanged particle, and 
\beq
R_{2\to 2} \sim \frac{cT^2}{p}\log\left(\frac{pT}{m_0^2}\right)
\,.
\eeq 
Hence the reaction rate is typically slower the larger $p$, with an approximate scaling $R_{2\to 2}\propto 1/p$. Since $H\sim T^2/M_{\rm Pl}$ in the radiation-dominated era, the solution to Boltzmann equation in Eq.~\eqref{Beqphi2} is approximately of the form
\beq\label{approxfphi}
f_\phi(p,T)/f_{\rm eq}\sim 1-\exp\left[-\frac{cM_{\rm Pl}}{p}\left(1+\log\frac{p}{T}\right)\right]\,.
\eeq 
At a given temperature, the density of $\phi$ particles drops faster below $f_{\rm eq}$ the larger the momentum. As a result, it takes more time for fast particles to reach equilibrium. \\

The approximate solution in Eq.~\eqref{approxfphi} proves useful for identifying which region(s) of the momentum distribution is relevant for DM production. The integrand of the collision term of the $\phi\phi\to \bar\chi\chi$ process involves the product $\eta\equiv f_\phi(p_1)f_\phi(p_2)$ where $p_{1,2}$ are the momenta of the two colliding $\phi$ particles. If $\phi$ were in equilibrium with the SM, assuming Maxwell-Boltzmann statisitics, the above product would only depend on the total momentum, $\eta_{\rm eq}=f_{\rm eq}(p_1)f_{\rm eq}(p_2)= \exp(-p_+)$ with $p_+\equiv p_1+p_2$, and all possible repartitions of $p_+$ between $p_1$ and $p_2$ would equally contribute. This is not necessarily the case for $\phi$ particles out-of-equilibrium. Consider the distribution of Eq.~\eqref{approxfphi} in evaluating $\eta$ in two distinct regimes where both particles have comparable momenta $p_1\sim p_2\sim p_+/2$, or one particle carries all the total required momentum, $p_1\sim p_+$ and $p_2\sim 0$. In the first regime, $\eta/\eta_{\rm eq}\sim\left[1-\exp(-2cM_{\rm Pl}/p_+)\right]^2\sim (2cM_{\rm Pl}/p_+)^2$, where  in the last expression we used $p_+\gg cM_{\rm Pl}$, typically valid out-of-equilibrium. Conversely, in the second regime the suppression is less severe, $\eta/\eta_{\rm eq}\sim 1-\exp(-c M_{\rm Pl}/p_+)\sim cM_{\rm Pl}/p_+$, because slow particles are close to equilibrium. Hence, production of heavy DM from much lighter, out-of-equilibrium $\phi$ fusion is dominated by fast particles colliding slow ones.

\section{Relevant formulae for direct detection} 
\label{app:dd}

We  parameterize the nuclear form-factor with a spherically-symmetric Fermi distribution, $F(q)=c\int r^2 dr e^{-iqr}/ [1+\exp[(r-R_A)/a]] $ where
 the parameter $R_A =(1.23A^{1/3} -0.6){\rm fm}$ ($A$ denoting the mass number of the nucleus) and the surface thickness $a=0.52 {\rm fm}$ have been extracted from muon scattering data for various nuclei ~\cite{Lewin:298578,Belanger:2008sj}. The normalization constant $c$ is such that  $F(q=0)=1$.
 
We assume the velocity distribution of DM is a Maxwellian centered on $v_0=220\,$km/sec, the galactic velocity. Further including the effects of both the velocity of the Earth relative to the galactic rest frame $v_E=232\,$km/sec and the galactic escape velocity $v_{\rm esc}=544\,$km/sec yields for the velocity distribution in Eq.~\ref{dRdER}~\cite{Lewin:298578}  
\beq
\eta(q^2)=\frac{k_0}{k_1}\left[\frac{\sqrt{\pi} v_0}{4v_E}\left({\rm erf}_+
-{\rm erf}_-
\right)-\delta \exp\left(-\frac{v_{\rm esc}^2}{v_0^2}\right)\right]\,,
\eeq  
where $k_1/k_0\equiv {\rm erf}(v_{\rm esc}/v_0)-2v_{\rm esc}/(\sqrt{\pi} v_0)\exp(-v_{\rm esc}^2/v_0^2)$, ${\rm erf}_\pm\equiv {\rm erf}(v_\pm/v_0)$, $\delta\equiv (v_+-v_-)/2v_E$ with $v_\pm = v_{\rm min}\pm v_E$ if $v_{\rm min}<v_{\rm esc}\mp v_E$ 
and $v_\pm = v_{\rm esc}$ otherwise, $v_{\rm min}=q/(2\mu_{\chi T})$ and ${\rm erf}$ denotes the error function.

\bibliographystyle{utphys}
\bibliography{DMrefs}

\providecommand{\href}[2]{#2}\begingroup\raggedright\begin{thebibliography}{10}

\bibitem{Alexander:2016aln}
J.~Alexander {\em et al.}, ``{Dark Sectors 2016 Workshop: Community Report},''
\newblock 2016.
\newblock \href{http://arxiv.org/abs/1608.08632}{{\tt arXiv:1608.08632
  [hep-ph]}}.
\newblock
\url{http://lss.fnal.gov/archive/2016/conf/fermilab-conf-16-421.pdf}.
\newblock

\bibitem{Battaglieri:2017aum}
M.~Battaglieri {\em et al.}, ``{US Cosmic Visions: New Ideas in Dark Matter
  2017: Community Report},'' in {\em {U.S. Cosmic Visions: New Ideas in Dark
  Matter College Park, MD, USA, March 23-25, 2017}}.
\newblock 2017.
\newblock \href{http://arxiv.org/abs/1707.04591}{{\tt arXiv:1707.04591
  [hep-ph]}}.
\newblock
\url{http://lss.fnal.gov/archive/2017/conf/fermilab-conf-17-282-ae-ppd-t.pdf}.
\newblock

\bibitem{Spergel:1999mh}
D.~N. Spergel and P.~J. Steinhardt, ``{Observational evidence for
  selfinteracting cold dark matter},''
  \href{http://dx.doi.org/10.1103/PhysRevLett.84.3760}{{\em Phys. Rev. Lett.}
  {\bf 84} (2000)  3760--3763},
\href{http://arxiv.org/abs/astro-ph/9909386}{{\tt arXiv:astro-ph/9909386
  [astro-ph]}}.

\bibitem{BoylanKolchin:2011de}
M.~Boylan-Kolchin, J.~S. Bullock, and M.~Kaplinghat, ``{Too big to fail? The
  puzzling darkness of massive Milky Way subhaloes},''
  \href{http://dx.doi.org/10.1111/j.1745-3933.2011.01074.x}{{\em Mon. Not. Roy.
  Astron. Soc.} {\bf 415} (2011)  L40},
\href{http://arxiv.org/abs/1103.0007}{{\tt arXiv:1103.0007 [astro-ph.CO]}}.

\bibitem{Oman:2015xda}
K.~A. Oman {\em et al.}, ``{The unexpected diversity of dwarf galaxy rotation
  curves},'' \href{http://dx.doi.org/10.1093/mnras/stv1504}{{\em Mon. Not. Roy.
  Astron. Soc.} {\bf 452} (2015) no.~4, 3650--3665},
\href{http://arxiv.org/abs/1504.01437}{{\tt arXiv:1504.01437 [astro-ph.GA]}}.

\bibitem{McDonald:2001vt}
J.~McDonald, ``{Thermally generated gauge singlet scalars as selfinteracting
  dark matter},'' \href{http://dx.doi.org/10.1103/PhysRevLett.88.091304}{{\em
  Phys. Rev. Lett.} {\bf 88} (2002)  091304},
\href{http://arxiv.org/abs/hep-ph/0106249}{{\tt arXiv:hep-ph/0106249
  [hep-ph]}}.

\bibitem{Hall:2009bx}
L.~J. Hall, K.~Jedamzik, J.~March-Russell, and S.~M. West, ``{Freeze-In
  Production of FIMP Dark Matter},''
  \href{http://dx.doi.org/10.1007/JHEP03(2010)080}{{\em JHEP} {\bf 03} (2010)
  080},
\href{http://arxiv.org/abs/0911.1120}{{\tt arXiv:0911.1120 [hep-ph]}}.

\bibitem{Bernal:2017kxu}
N.~Bernal, M.~Heikinheimo, T.~Tenkanen, K.~Tuominen, and V.~Vaskonen, ``{The
  Dawn of FIMP Dark Matter: A Review of Models and Constraints},''
  \href{http://dx.doi.org/10.1142/S0217751X1730023X}{{\em Int. J. Mod. Phys.}
  {\bf A32} (2017) no.~27, 1730023},
\href{http://arxiv.org/abs/1706.07442}{{\tt arXiv:1706.07442 [hep-ph]}}.

\bibitem{Hambye:2018dpi}
T.~Hambye, M.~H.~G. Tytgat, J.~Vandecasteele, and L.~Vanderheyden, ``{Dark
  matter direct detection is testing freeze-in},''
  \href{http://dx.doi.org/10.1103/PhysRevD.98.075017}{{\em Phys. Rev.} {\bf
  D98} (2018) no.~7, 075017},
\href{http://arxiv.org/abs/1807.05022}{{\tt arXiv:1807.05022 [hep-ph]}}.

\bibitem{Essig:2017kqs}
R.~Essig, T.~Volansky, and T.-T. Yu, ``{New Constraints and Prospects for
  sub-GeV Dark Matter Scattering off Electrons in Xenon},''
  \href{http://dx.doi.org/10.1103/PhysRevD.96.043017}{{\em Phys. Rev.} {\bf
  D96} (2017) no.~4, 043017},
\href{http://arxiv.org/abs/1703.00910}{{\tt arXiv:1703.00910 [hep-ph]}}.

\bibitem{Arcadi:2013aba}
G.~Arcadi and L.~Covi, ``{Minimal Decaying Dark Matter and the LHC},''
  \href{http://dx.doi.org/10.1088/1475-7516/2013/08/005}{{\em JCAP} {\bf 1308}
  (2013)  005},
\href{http://arxiv.org/abs/1305.6587}{{\tt arXiv:1305.6587 [hep-ph]}}.

\bibitem{Roland:2015yoa}
S.~B. Roland, B.~Shakya, and J.~D. Wells, ``{PeV neutrinos and a 3.5 keV x-ray
  line from a PeV-scale supersymmetric neutrino sector},''
  \href{http://dx.doi.org/10.1103/PhysRevD.92.095018}{{\em Phys. Rev.} {\bf
  D92} (2015) no.~9, 095018},
\href{http://arxiv.org/abs/1506.08195}{{\tt arXiv:1506.08195 [hep-ph]}}.

\bibitem{Fradette:2014sza}
A.~Fradette, M.~Pospelov, J.~Pradler, and A.~Ritz, ``{Cosmological Constraints
  on Very Dark Photons},''
  \href{http://dx.doi.org/10.1103/PhysRevD.90.035022}{{\em Phys. Rev.} {\bf
  D90} (2014) no.~3, 035022},
\href{http://arxiv.org/abs/1407.0993}{{\tt arXiv:1407.0993 [hep-ph]}}.

\bibitem{Bernal:2015ova}
N.~Bernal, X.~Chu, C.~Garcia-Cely, T.~Hambye, and B.~Zaldivar, ``{Production
  Regimes for Self-Interacting Dark Matter},''
  \href{http://dx.doi.org/10.1088/1475-7516/2016/03/018}{{\em JCAP} {\bf 03}
  (2016)  018}, \href{http://arxiv.org/abs/1510.08063}{{\tt arXiv:1510.08063
  [hep-ph]}}.

\bibitem{Berger:2016vxi}
J.~Berger, K.~Jedamzik, and D.~G.~E. Walker, ``{Cosmological Constraints on
  Decoupled Dark Photons and Dark Higgs},''
  \href{http://dx.doi.org/10.1088/1475-7516/2016/11/032}{{\em JCAP} {\bf 1611}
  (2016)  032},
\href{http://arxiv.org/abs/1605.07195}{{\tt arXiv:1605.07195 [hep-ph]}}.

\bibitem{Co:2015pka}
R.~T. Co, F.~D'Eramo, L.~J. Hall, and D.~Pappadopulo, ``{Freeze-In Dark Matter
  with Displaced Signatures at Colliders},''
  \href{http://dx.doi.org/10.1088/1475-7516/2015/12/024}{{\em JCAP} {\bf 1512}
  (2015) no.~12, 024},
\href{http://arxiv.org/abs/1506.07532}{{\tt arXiv:1506.07532 [hep-ph]}}.

\bibitem{Evans:2016zau}
J.~A. Evans and J.~Shelton, ``{Long-Lived Staus and Displaced Leptons at the
  LHC},'' \href{http://dx.doi.org/10.1007/JHEP04(2016)056}{{\em JHEP} {\bf 04}
  (2016)  056},
\href{http://arxiv.org/abs/1601.01326}{{\tt arXiv:1601.01326 [hep-ph]}}.

\bibitem{Hessler:2016kwm}
A.~G. Hessler, A.~Ibarra, E.~Molinaro, and S.~Vogl, ``{Probing the scotogenic
  FIMP at the LHC},'' \href{http://dx.doi.org/10.1007/JHEP01(2017)100}{{\em
  JHEP} {\bf 01} (2017)  100},
\href{http://arxiv.org/abs/1611.09540}{{\tt arXiv:1611.09540 [hep-ph]}}.

\bibitem{Ghosh:2017vhe}
A.~Ghosh, T.~Mondal, and B.~Mukhopadhyaya, ``{Heavy stable charged tracks as
  signatures of non-thermal dark matter at the LHC : a study in some
  non-supersymmetric scenarios},''
  \href{http://dx.doi.org/10.1007/JHEP12(2017)136}{{\em JHEP} {\bf 12} (2017)
  136},
\href{http://arxiv.org/abs/1706.06815}{{\tt arXiv:1706.06815 [hep-ph]}}.

\bibitem{Calibbi:2018fqf}
L.~Calibbi, L.~Lopez-Honorez, S.~Lowette, and A.~Mariotti, ``{Singlet-Doublet
  Dark Matter Freeze-in: LHC displaced signatures versus cosmology},''
  \href{http://dx.doi.org/10.1007/JHEP09(2018)037}{{\em JHEP} {\bf 09} (2018)
  037},
\href{http://arxiv.org/abs/1805.04423}{{\tt arXiv:1805.04423 [hep-ph]}}.

\bibitem{Belanger:2018sti}
G.~B\'elanger {\em et al.}, ``{LHC-friendly minimal freeze-in models},''
  \href{http://dx.doi.org/10.1007/JHEP02(2019)186}{{\em JHEP} {\bf 02} (2019)
  186},
\href{http://arxiv.org/abs/1811.05478}{{\tt arXiv:1811.05478 [hep-ph]}}.

\bibitem{Alimena:2019zri}
J.~Alimena {\em et al.}, ``{Searching for long-lived particles beyond the
  Standard Model at the Large Hadron Collider},''
\href{http://arxiv.org/abs/1903.04497}{{\tt arXiv:1903.04497 [hep-ex]}}.

\bibitem{Curtin:2018mvb}
D.~Curtin {\em et al.}, ``{Long-Lived Particles at the Energy Frontier: The
  MATHUSLA Physics Case},''
\href{http://arxiv.org/abs/1806.07396}{{\tt arXiv:1806.07396 [hep-ph]}}.

\bibitem{Ariga:2018uku}
{\bf FASER} Collaboration, A.~Ariga {\em et al.}, ``{FASER?s physics reach for
  long-lived particles},''
  \href{http://dx.doi.org/10.1103/PhysRevD.99.095011}{{\em Phys. Rev.} {\bf
  D99} (2019) no.~9, 095011},
\href{http://arxiv.org/abs/1811.12522}{{\tt arXiv:1811.12522 [hep-ph]}}.

\bibitem{No:2019gvl}
J.~M. No, P.~Tunney, and B.~Zaldivar, ``{Probing Dark Matter freeze-in with
  long-lived particle signatures: MATHUSLA, HL-LHC and FCC-hh},''
\href{http://arxiv.org/abs/1908.11387}{{\tt arXiv:1908.11387 [hep-ph]}}.

\bibitem{Heeba:2019jho}
S.~Heeba and F.~Kahlhoefer, ``{Probing the freeze-in mechanism in dark matter
  models with $U(1)^\prime$ gauge extensions},''
\href{http://arxiv.org/abs/1908.09834}{{\tt arXiv:1908.09834 [hep-ph]}}.

\bibitem{Boehm:2003hm}
C.~Boehm and P.~Fayet, ``{Scalar dark matter candidates},''
  \href{http://dx.doi.org/10.1016/j.nuclphysb.2004.01.015}{{\em Nucl. Phys.}
  {\bf B683} (2004)  219--263},
\href{http://arxiv.org/abs/hep-ph/0305261}{{\tt arXiv:hep-ph/0305261
  [hep-ph]}}.

\bibitem{Zurek-lightDM}
S.~Knapen, T.~Lin, and K.~M. Zurek, ``{Light Dark Matter: Models and
  Constraints},'' \href{http://dx.doi.org/10.1103/PhysRevD.96.115021}{{\em
  Phys. Rev.} {\bf D96} (2017) no.~11, 115021},
\href{http://arxiv.org/abs/1709.07882}{{\tt arXiv:1709.07882 [hep-ph]}}.

\bibitem{Buckley:2009in}
M.~R. Buckley and P.~J. Fox, ``{Dark Matter Self-Interactions and Light Force
  Carriers},'' \href{http://dx.doi.org/10.1103/PhysRevD.81.083522}{{\em Phys.
  Rev.} {\bf D81} (2010)  083522},
\href{http://arxiv.org/abs/0911.3898}{{\tt arXiv:0911.3898 [hep-ph]}}.

\bibitem{Kaplinghat:2015aga}
M.~Kaplinghat, S.~Tulin, and H.-B. Yu, ``{Dark Matter Halos as Particle
  Colliders: Unified Solution to Small-Scale Structure Puzzles from Dwarfs to
  Clusters},'' \href{http://dx.doi.org/10.1103/PhysRevLett.116.041302}{{\em
  Phys. Rev. Lett.} {\bf 116} (2016) no.~4, 041302},
\href{http://arxiv.org/abs/1508.03339}{{\tt arXiv:1508.03339 [astro-ph.CO]}}.

\bibitem{Kim:2008pp}
Y.~G. Kim, K.~Y. Lee, and S.~Shin, ``{Singlet fermionic dark matter},''
  \href{http://dx.doi.org/10.1088/1126-6708/2008/05/100}{{\em JHEP} {\bf 05}
  (2008)  100}, \href{http://arxiv.org/abs/0803.2932}{{\tt arXiv:0803.2932
  [hep-ph]}}.

\bibitem{LopezHonorez:2012kv}
L.~Lopez-Honorez, T.~Schwetz, and J.~Zupan, ``{Higgs portal, fermionic dark
  matter, and a Standard Model like Higgs at 125 GeV},''
  \href{http://dx.doi.org/10.1016/j.physletb.2012.07.017}{{\em Phys.\ Lett.\ B}
  {\bf 716} (2012)  179--185}, \href{http://arxiv.org/abs/1203.2064}{{\tt
  arXiv:1203.2064 [hep-ph]}}.

\bibitem{Krnjaic:2015mbs}
G.~Krnjaic, ``{Probing Light Thermal Dark-Matter With a Higgs Portal
  Mediator},'' \href{http://dx.doi.org/10.1103/PhysRevD.94.073009}{{\em Phys.
  Rev.} {\bf D94} (2016) no.~7, 073009},
\href{http://arxiv.org/abs/1512.04119}{{\tt arXiv:1512.04119 [hep-ph]}}.

\bibitem{Hambye:2019dwd}
T.~Hambye, M.~H.~G. Tytgat, J.~Vandecasteele, and L.~Vanderheyden, ``{Dark
  matter from dark photons: a taxonomy of dark matter production},''
  \href{http://dx.doi.org/10.1103/PhysRevD.100.095018}{{\em Phys. Rev.} {\bf
  D100} (2019) no.~9, 095018},
\href{http://arxiv.org/abs/1908.09864}{{\tt arXiv:1908.09864 [hep-ph]}}.

\bibitem{Dvorkin:2019zdi}
C.~Dvorkin, T.~Lin, and K.~Schutz, ``{Making dark matter out of light:
  freeze-in from plasma effects},''
\href{http://arxiv.org/abs/1902.08623}{{\tt arXiv:1902.08623 [hep-ph]}}.

\bibitem{Belanger:2018mqt}
G.~B\'elanger, F.~Boudjema, A.~Goudelis, A.~Pukhov, and B.~Zaldivar,
  ``{micrOMEGAs5.0 : Freeze-in},''
  \href{http://dx.doi.org/10.1016/j.cpc.2018.04.027}{{\em Comput. Phys.
  Commun.} {\bf 231} (2018)  173--186},
\href{http://arxiv.org/abs/1801.03509}{{\tt arXiv:1801.03509 [hep-ph]}}.

\bibitem{Batell:2018fqo}
B.~Batell, A.~Freitas, A.~Ismail, and D.~Mckeen, ``{Probing Light Dark Matter
  with a Hadrophilic Scalar Mediator},''
  \href{http://dx.doi.org/10.1103/PhysRevD.100.095020}{{\em Phys. Rev.} {\bf
  D100} (2019) no.~9, 095020},
\href{http://arxiv.org/abs/1812.05103}{{\tt arXiv:1812.05103 [hep-ph]}}.

\bibitem{Batell:2017kty}
B.~Batell, A.~Freitas, A.~Ismail, and D.~Mckeen, ``{Flavor-specific scalar
  mediators},'' \href{http://dx.doi.org/10.1103/PhysRevD.98.055026}{{\em Phys.
  Rev.} {\bf D98} (2018) no.~5, 055026},
\href{http://arxiv.org/abs/1712.10022}{{\tt arXiv:1712.10022 [hep-ph]}}.

\bibitem{Egana-Ugrinovic:2018znw}
D.~Egana-Ugrinovic, S.~Homiller, and P.~Meade, ``{Aligned and Spontaneous
  Flavor Violation},''
\href{http://arxiv.org/abs/1811.00017}{{\tt arXiv:1811.00017 [hep-ph]}}.

\bibitem{Shifman:1978zn}
M.~A. Shifman, A.~I. Vainshtein, and V.~I. Zakharov, ``{Remarks on Higgs Boson
  Interactions with Nucleons},''
\href{http://dx.doi.org/10.1016/0370-2693(78)90481-1}{{\em Phys. Lett.} {\bf
  78B} (1978)  443--446}.

\bibitem{Bhattacharya:2016zcn}
T.~Bhattacharya, V.~Cirigliano, S.~Cohen, R.~Gupta, H.-W. Lin, and B.~Yoon,
  ``{Axial, Scalar and Tensor Charges of the Nucleon from 2+1+1-flavor Lattice
  QCD},'' \href{http://dx.doi.org/10.1103/PhysRevD.94.054508}{{\em Phys. Rev.}
  {\bf D94} (2016) no.~5, 054508},
\href{http://arxiv.org/abs/1606.07049}{{\tt arXiv:1606.07049 [hep-lat]}}.

\bibitem{Donoghue:1990xh}
J.~F. Donoghue, J.~Gasser, and H.~Leutwyler, ``{The Decay of a Light Higgs
  Boson},''
\href{http://dx.doi.org/10.1016/0550-3213(90)90474-R}{{\em Nucl. Phys.} {\bf
  B343} (1990)  341--368}.

\bibitem{Bijnens:1998fm}
J.~Bijnens, G.~Colangelo, and P.~Talavera, ``{The Vector and scalar
  form-factors of the pion to two loops},''
  \href{http://dx.doi.org/10.1088/1126-6708/1998/05/014}{{\em JHEP} {\bf 05}
  (1998)  014},
\href{http://arxiv.org/abs/hep-ph/9805389}{{\tt arXiv:hep-ph/9805389
  [hep-ph]}}.

\bibitem{Bellac:2011kqa}
M.~L. Bellac, \href{http://dx.doi.org/10.1017/CBO9780511721700}{{\em {Thermal
  Field Theory}}}.
\newblock Cambridge Monographs on Mathematical Physics. Cambridge University
  Press, 3, 2011.

\bibitem{Kapusta:1989tk}
J.~I. Kapusta, {\em {Finite Temperature Field Theory}}, vol.~360 of {\em
  Cambridge Monographs on Mathematical Physics}.
\newblock Cambridge University Press, Cambridge, 1989.

\bibitem{Su:2011zv}
N.~Su, ``{A Gauge-Invariant Reorganization of Thermal Gauge Theory},'' other
  thesis, 4, 2011.

\bibitem{Braaten:1989mz}
E.~Braaten and R.~D. Pisarski, ``{Soft Amplitudes in Hot Gauge Theories: A
  General Analysis},''
\href{http://dx.doi.org/10.1016/0550-3213(90)90508-B}{{\em Nucl. Phys.} {\bf
  B337} (1990)  569--634}.

\bibitem{tsytovich1961spatial}
V.~Tsytovich, ``Spatial dispersion in a relativistic plasma,'' {\em Sov. Phys.
  JETP} {\bf 13} (1961) no.~6, 1249.

\bibitem{Weldon:1982aq}
H.~A. Weldon, ``{Covariant Calculations at Finite Temperature: The Relativistic
  Plasma},''
\href{http://dx.doi.org/10.1103/PhysRevD.26.1394}{{\em Phys. Rev.} {\bf D26}
  (1982)  1394}.

\bibitem{Weldon:1989ys}
H.~Weldon, ``{Dynamical Holes in the Quark - Gluon Plasma},''
  \href{http://dx.doi.org/10.1103/PhysRevD.40.2410}{{\em Phys. Rev. D} {\bf 40}
  (1989)  2410}.

\bibitem{Giudice:2003jh}
G.~F. Giudice, A.~Notari, M.~Raidal, A.~Riotto, and A.~Strumia, ``{Towards a
  complete theory of thermal leptogenesis in the SM and MSSM},''
  \href{http://dx.doi.org/10.1016/j.nuclphysb.2004.02.019}{{\em Nucl. Phys.}
  {\bf B685} (2004)  89--149},
\href{http://arxiv.org/abs/hep-ph/0310123}{{\tt arXiv:hep-ph/0310123
  [hep-ph]}}.

\bibitem{Weldon:1982bn}
H.~A. Weldon, ``{Effective Fermion Masses of Order gT in High Temperature Gauge
  Theories with Exact Chiral Invariance},''
\href{http://dx.doi.org/10.1103/PhysRevD.26.2789}{{\em Phys. Rev.} {\bf D26}
  (1982)  2789}.

\bibitem{Planck2018}
{\bf Planck} Collaboration, N.~Aghanim {\em et al.}, ``{Planck 2018 results.
  VI. Cosmological parameters},''
\href{http://arxiv.org/abs/1807.06209}{{\tt arXiv:1807.06209 [astro-ph.CO]}}.

\bibitem{Aprile:2018dbl}
{\bf XENON} Collaboration, E.~Aprile {\em et al.}, ``{Dark Matter Search
  Results from a One Ton-Year Exposure of XENON1T},''
  \href{http://dx.doi.org/10.1103/PhysRevLett.121.111302}{{\em Phys. Rev.
  Lett.} {\bf 121} (2018) no.~11, 111302},
\href{http://arxiv.org/abs/1805.12562}{{\tt arXiv:1805.12562 [astro-ph.CO]}}.

\bibitem{WittenDM}
M.~W. Goodman and E.~Witten, ``{Detectability of Certain Dark Matter
  Candidates},'' \href{http://dx.doi.org/10.1103/PhysRevD.31.3059}{{\em Phys.
  Rev.} {\bf D31} (1985)  3059}.
[,325(1984)].

\bibitem{Lewin:298578}
J.~D. Lewin and P.~F. Smith, ``{Review of mathematics, numerical factors, and
  corrections for dark matter experiments based on elastic nuclear recoil},''
  Tech. Rep. RAL-TR-95-024, RAL, Chilton, Feb, 1996.
\newblock \url{http://cds.cern.ch/record/298578}.

\bibitem{Belanger:2020gnr}
G.~Belanger, A.~Mjallal, and A.~Pukhov, ``{Recasting direct detection limits
  within micrOMEGAs and implication for non-standard Dark Matter scenarios},''
  \href{http://arxiv.org/abs/2003.08621}{{\tt arXiv:2003.08621 [hep-ph]}}.

\bibitem{Agnes:2018ves}
{\bf DarkSide} Collaboration, P.~Agnes {\em et al.}, ``{Low-Mass Dark Matter
  Search with the DarkSide-50 Experiment},''
  \href{http://dx.doi.org/10.1103/PhysRevLett.121.081307}{{\em Phys. Rev.
  Lett.} {\bf 121} (2018) no.~8, 081307},
\href{http://arxiv.org/abs/1802.06994}{{\tt arXiv:1802.06994 [astro-ph.HE]}}.

\bibitem{Aprile:2019xxb}
{\bf XENON} Collaboration, E.~Aprile {\em et al.}, ``{Light Dark Matter Search
  with Ionization Signals in XENON1T},''
  \href{http://dx.doi.org/10.1103/PhysRevLett.123.251801}{{\em Phys.\ Rev.\
  Lett.} {\bf 123} (2019) no.~25, 251801},
  \href{http://arxiv.org/abs/1907.11485}{{\tt arXiv:1907.11485 [hep-ex]}}.

\bibitem{Aprile:2019jmx}
{\bf XENON} Collaboration, E.~Aprile {\em et al.}, ``{Search for Light Dark
  Matter Interactions Enhanced by the Migdal Effect or Bremsstrahlung in
  XENON1T},'' \href{http://dx.doi.org/10.1103/PhysRevLett.123.241803}{{\em
  Phys.\ Rev.\ Lett.} {\bf 123} (2019) no.~24, 241803},
  \href{http://arxiv.org/abs/1907.12771}{{\tt arXiv:1907.12771 [hep-ex]}}.

\bibitem{Agnese:2016cpb}
{\bf SuperCDMS} Collaboration, R.~Agnese {\em et al.}, ``{Projected Sensitivity
  of the SuperCDMS SNOLAB experiment},''
  \href{http://dx.doi.org/10.1103/PhysRevD.95.082002}{{\em Phys. Rev.} {\bf
  D95} (2017) no.~8, 082002},
\href{http://arxiv.org/abs/1610.00006}{{\tt arXiv:1610.00006
  [physics.ins-det]}}.

\bibitem{Frugiuele:2016rii}
C.~Frugiuele, E.~Fuchs, G.~Perez, and M.~Schlaffer, ``{Constraining New Physics
  Models with Isotope Shift Spectroscopy},''
  \href{http://dx.doi.org/10.1103/PhysRevD.96.015011}{{\em Phys. Rev.} {\bf
  D96} (2017) no.~1, 015011},
\href{http://arxiv.org/abs/1602.04822}{{\tt arXiv:1602.04822 [hep-ph]}}.

\bibitem{PhysRevLett.68.1472}
H.~Leeb and J.~Schmiedmayer, ``{Constraint on hypothetical light interacting
  bosons from low-energy neutron experiments},''
  \href{http://dx.doi.org/10.1103/PhysRevLett.68.1472}{{\em Phys. Rev. Lett.}
  {\bf 68} (1992)  1472--1475}.

\bibitem{Willey:1982mc}
R.~S. Willey and H.~L. Yu, ``{Neutral Higgs Boson From Decays of Heavy Flavored
  Mesons},''
\href{http://dx.doi.org/10.1103/PhysRevD.26.3086}{{\em Phys. Rev.} {\bf D26}
  (1982)  3086}.

\bibitem{Lees:2013kla}
{\bf BaBar} Collaboration, J.~P. Lees {\em et al.}, ``{Search for $B \to
  K^{(*)} \nu \overline \nu$ and invisible quarkonium decays},''
  \href{http://dx.doi.org/10.1103/PhysRevD.87.112005}{{\em Phys. Rev.} {\bf
  D87} (2013) no.~11, 112005},
\href{http://arxiv.org/abs/1303.7465}{{\tt arXiv:1303.7465 [hep-ex]}}.

\bibitem{Kpinunu}
{\bf E949} Collaboration, A.~V. Artamonov {\em et al.}, ``{New measurement of
  the $K^{+} \to \pi^{+} \nu \bar{\nu}$ branching ratio},''
  \href{http://dx.doi.org/10.1103/PhysRevLett.101.191802}{{\em Phys. Rev.
  Lett.} {\bf 101} (2008)  191802},
\href{http://arxiv.org/abs/0808.2459}{{\tt arXiv:0808.2459 [hep-ex]}}.

\bibitem{NA62Kpinunu}
{\bf NA62} Collaboration, S.~Martellotti, ``{$K^+\rightarrow \pi^+\nu\bar\nu$
  decay and NP searches at NA62},''
\href{http://arxiv.org/abs/1807.08340}{{\tt arXiv:1807.08340 [hep-ex]}}.

\bibitem{Bergsma:1985qz}
{\bf CHARM} Collaboration, F.~Bergsma {\em et al.}, ``{Search for Axion Like
  Particle Production in 400-{GeV} Proton - Copper Interactions},''
  \href{http://dx.doi.org/10.1016/0370-2693(85)90400-9}{{\em Phys.\ Lett.\ B}
  {\bf 157} (1985)  458--462}.

\bibitem{Raffelt:1996wa}
G.~G. Raffelt, {\em {Stars as laboratories for fundamental physics}}.
\newblock 1996.
\newblock
\url{https://wwwth.mpp.mpg.de/members/raffelt/mypapers/199613.pdf}.
\newblock

\bibitem{Arbey:2018zfh}
A.~Arbey, J.~Auffinger, K.~Hickerson, and E.~Jenssen, ``{AlterBBN v2: A public
  code for calculating Big-Bang nucleosynthesis constraints in alternative
  cosmologies},'' \href{http://dx.doi.org/10.1016/j.cpc.2019.106982}{{\em
  Comput.\ Phys.\ Commun.} {\bf 248} (2020)  106982},
  \href{http://arxiv.org/abs/1806.11095}{{\tt arXiv:1806.11095 [astro-ph.CO]}}.

\bibitem{Belanger:2008sj}
G.~Belanger, F.~Boudjema, A.~Pukhov, and A.~Semenov, ``{Dark matter direct
  detection rate in a generic model with micrOMEGAs 2.2},''
  \href{http://dx.doi.org/10.1016/j.cpc.2008.11.019}{{\em Comput. Phys.
  Commun.} {\bf 180} (2009)  747--767},
\href{http://arxiv.org/abs/0803.2360}{{\tt arXiv:0803.2360 [hep-ph]}}.

\end{thebibliography}\endgroup

\end{document}